%% file: main.tex
\documentclass[conference]{IEEEtran}
\usepackage{cite}
\usepackage{amsmath,amssymb,amsfonts}
\usepackage{booktabs}
\usepackage{tikz}
\usepackage{url}
\usepackage{graphicx}

\newcommand{\aname}{\texttt{DARE}}
\newcommand{\anameexpansion}{Difficulty Analysis for Repair Enhancement}
\newcommand{\anamefull}{\aname{} (\anameexpansion)}

\usepackage{tcolorbox}
\makeatletter
\newcommand{\mybox}[1]{%
	\setbox0=\hbox{#1}%
	\setlength{\@tempdima}{\dimexpr\wd0+13pt}%
	\begin{tcolorbox}[boxrule=0.5pt, colback=white, arc=4pt,
		left=6pt,right=6pt,top=6pt,bottom=6pt,boxsep=0pt,]
		#1
	\end{tcolorbox}
}

\newtcolorbox{boxA}{
    boxrule = 1pt,
    colframe = black 
}

\begin{document}
\title{\aname: \anameexpansion}

\author{

    \IEEEauthorblockN{Jikwang Lee, Seungwook Jung, Jaechang Nam}
    \IEEEauthorblockA{
    \textit{Handong Global University}, Pohang, South Korea\\
    lucas20606@gmail.com, wjdtmddnr123@gmail.com, jcnam@handong.edu}
}

\maketitle
\input{sections/abstract}

\input{sections/introduction}
\input{sections/background}
\input{sections/approach}
\input{sections/experimental_setup}

\input{sections/results}
\input{sections/discussion}
\input{sections/conclusion}

\bibliographystyle{IEEEtran}
\bibliography{ref}
\end{document}

%% file: sections/abstract.tex
\begin{abstract}
LLM-based coding agents are primarily evaluated by their solve rates. However, this metric does not reveal how difficult the solved tasks are or account for differences in task difficulty across benchmarks. Determining the difficulty of individual tasks by repeatedly running multiple agents is costly, while a predicted difficulty score alone provides limited guidance on what support an agent needs. To address these challenges, we propose \anamefull, which integrates repair-task difficulty measurement, prediction, feature-group contribution analysis, and task-specific repair support. We define continuous task difficulty as the observed failure rate across multiple coding agents. The \aname{} difficulty model predicts this difficulty using 151 static features extracted from benchmark artifacts. The framework also analyzes the contributions of seven Repair-Task Feature Groups to identify the characteristics most salient to each task and uses this information to select task-specific agent support. We evaluate \aname{} on four benchmark partitions comprising 2,594 tasks. The difficulty model's aligned predictions achieve Spearman's $\rho = .476$--$.621$ on held-out partitions, and the most salient Feature Groups vary across tasks and benchmarks. Furthermore, coding-agent experiments show that matching support to each task’s salient Feature Group raises repair success from 54.7\% to 79.6\% on nine high-difficulty tasks. These findings demonstrate that difficulty prediction and interpretation can lead to practical improvements in automated program repair (APR), while also providing a basis for difficulty-stratified agent evaluation, cross-benchmark comparisons, and identifying areas of repair capability that warrant further improvement.
\end{abstract}

%% file: sections/introduction.tex
\section{Introduction}
\label{sec:introduction}

Evaluating the performance of LLM-based coding agents solely by their overall solve rate can obscure both their ability to solve difficult tasks and how that ability has improved. Coding agents have shown rapid improvements in resolving real-world software issues. For example, reported solve rates on SWE-bench Verified increased from 33.2\% in 2024 to 80.9\% by early 2026~\cite{chowdhury2024verified,openai2026verifiedlimitations}. However, the overall solve rate alone cannot reveal whether these gains reflect progress on previously unsolved difficult tasks or improvements primarily on easier ones. Moreover, two agents with the same solve rate may solve tasks of different difficulty, and differences in solve rates across benchmarks may reflect differences in task-difficulty distributions as well as agent capabilities. Therefore, understanding agent performance and progress in automated program repair (APR) requires considering not only \emph{how many tasks agents solve}, but also \emph{how difficult those tasks are}.

Knowing the difficulty of individual tasks would enable more informative agent evaluation, but measuring it directly is expensive. Estimating task difficulty requires evaluating each task with multiple agents or conducting substantial manual assessments. A more efficient alternative is to predict difficulty from artifacts already available in a benchmark. Such predictions can support difficulty-stratified performance analysis, comparisons of difficulty distributions across benchmarks, and the selection of evaluation tasks without running a large number of agents on every task.

Prior studies have explored predicting task success or difficulty using task- and agent-related characteristics. However, how to leverage the task characteristics contributing to difficulty predictions to improve actual repair performance remains underexplored~\cite{alhaque2026difficulty,krsteski2026predicting,ge2026agentpsychometrics}.

To go beyond difficulty prediction, we propose \anamefull, a framework that combines difficulty prediction and feature-group contribution analysis to guide task-specific repair support. We define task difficulty as the observed failure rate across multiple coding agents. The \aname{} difficulty model uses a generalized additive model (GAM)~\cite{hastie1986gam} to estimate task difficulty from 151 static features extracted from benchmark artifacts. These features are organized into seven Repair-Task Feature Groups that capture different characteristics relevant to software repair. \aname{} quantifies each group's contribution to the model's difficulty prediction, identifies the most salient Feature Group for each task, and uses this information to guide the selection of corresponding repair support.

We evaluate \aname{} on four benchmark partitions comprising 2,594 tasks through three research questions. \textbf{RQ1:} How accurately does \aname{} predict repair difficulty? \textbf{RQ2:} How do Feature Group contributions vary with predicted task difficulty? \textbf{RQ3:} Does Feature-Group-guided support improve coding-agent repair success? The difficulty model generalizes to held-out partitions (Spearman $\rho=.476$--$.621$), and the most salient Feature Group varies across tasks. Matching support to this group raises a coding agent's repair success from 54.7\% to 79.6\% on nine high-difficulty tasks ($p=.0039$).

\noindent\textbf{Our main contributions are as follows:}
\begin{itemize}
  \item \textbf{Difficulty prediction across benchmarks.} We define repair-task difficulty using observed failure rates across multiple agents and evaluate its prediction on held-out benchmark partitions. We also use shared-agent scale alignment to adjust for differing agent populations.

  \item \textbf{Interpretable analysis of repair-task difficulty.} We organize 151 static features into seven Repair-Task Feature Groups reflecting different characteristics of software repair and analyze their task-level contributions to difficulty predictions. This reveals differences between tasks that a single difficulty score cannot capture.

  \item \textbf{Analysis-guided repair support.} We use each task's most salient Feature Group to guide the selection of task-specific support and evaluate its impact on coding-agent repair success, testing whether model-based difficulty interpretation is useful for practical repair assistance.
\end{itemize}

%% file: sections/background.tex
\section{Background and Related Work}
\label{sec:background}

\subsection{Automated Program Repair and Coding-Agent Evaluation}

Automated program repair (APR) aims to fix software defects automatically~\cite{gazzola2019survey}. With LLMs, repair now includes coding agents that explore repositories, modify code, and validate changes to resolve real-world issues~\cite{yang2024sweagent,zhang2024autocoderover}.

SWE-bench evaluates whether agent-generated patches resolve real GitHub issues~\cite{jimenez2024swebench}. SWE-bench Verified provides human-reviewed tasks~\cite{chowdhury2024verified}, while SWE-bench-Live adds newer issues for evaluation over time~\cite{zhang2025swebenchgoeslive}. Section~\ref{sec:benchmarks} describes the benchmark partitions used in this study.

Agent performance is commonly summarized by solve rate, the proportion of tasks resolved successfully. However, two agents with the same solve rate need not solve the same tasks. Aggregate solve rates alone therefore provide limited insight into the difficulty of individual tasks or the relationship between task characteristics and repair outcomes~\cite{alhaque2026difficulty,krsteski2026predicting,ge2026agentpsychometrics}. This motivates measuring and predicting task-level difficulty as a complement to overall agent evaluation.

\subsection{Repair-Task Difficulty Prediction}

Al-Haque and Johnson~\cite{alhaque2026difficulty} analyzed repeated runs of a single Qwen3-Coder--OpenHands configuration on CoderForge-Preview using 54 static features from issues, reference patches, and repositories. They reported AUC .863 for predicting any-success (whether a task was solved at least once) and also examined continuous pass-rate prediction and SHAP-based attribution.
Their single-agent, single-dataset setup does not show generalization across agent populations or benchmarks.

Krsteski and Meyer~\cite{krsteski2026predicting} predicted the relative ordering of benchmark-standardized item response theory (IRT) difficulty across 17 agentic benchmarks without running agents on target tasks. They showed that response-level AUC can reach .715 even with identical task-difficulty estimates, solely because of agent-ability differences. 
Thus, response-level AUC alone cannot show that a model captures task-difficulty differences.

Ge et al.~\cite{ge2026agentpsychometrics} combined static task information and agent characteristics in an IRT-based model to estimate task difficulty and agent ability and to predict success probabilities for unobserved task--agent pairs. They evaluated generalization to unseen benchmarks and used their predictions to select evaluation tasks efficiently. Their primary evaluation, however, focused on agent--task success/failure discrimination using AUC-ROC, rather than directly assessing the ranking and absolute error of task-level multi-agent failure rates or analyzing contributions from Repair-Task Feature Groups~\cite{ge2026agentpsychometrics}.

\aname{} is therefore distinguished not simply by predicting task outcomes from static features, but by three aspects of its design. First, it defines continuous repair-task difficulty using the observed failure rates of multiple coding agents. Second, it evaluates both difficulty ordering and absolute prediction error on held-out benchmark partitions. Third, it aggregates contributions from the same additive prediction model into seven Repair-Task Feature Groups to identify candidates for task-specific support (Sections~\ref{sec:repair_task_difficulty}--\ref{sec:dare} and~\ref{sec:evaluation_metrics}; RQ1--RQ2). RQ3 then evaluates whether support guided by these contributions improves coding-agent repair success. Thus, \aname{} connects difficulty prediction and Feature Group interpretation to a support stage whose practical effectiveness is evaluated through coding-agent experiments.

\subsection{Failure Analysis and Repair-Specific Support}

Liu et al.~\cite{liu2025failures} analyzed failures in unresolved issues and evaluated an improvement architecture informed by their findings.
Ceka et al.~\cite{ceka2026traceability}  examined agent execution traces to identify testing and debugging bottlenecks and discussed design implications.
Both identify recurring problems from outcomes or trajectories observed after execution~\cite{liu2025failures,ceka2026traceability}.

Other systems improve repair during execution: AutoCodeRover~\cite{zhang2024autocoderover} supports code search and fault localization; SpecRover~\cite{ruan2025specrover} uses specification inference and a reviewer agent; and RepairAgent~\cite{bouzenia2025repairagent} selects tools based on information and feedback gathered during execution.
These approaches focus on the repair process or its tool use~\cite{zhang2024autocoderover,ruan2025specrover,bouzenia2025repairagent}.

Unlike post-hoc analysis or feedback-driven tool selection, \aname{} identifies each task's most salient Feature Group from static benchmark artifacts before the target agent attempts repair (RQ2), selects corresponding assistance, and evaluates its impact (RQ3). 
Reference-patch-derived features may indirectly inform support selection, but the reference patches are not provided to the agent; Section~\ref{sec:threats-to-validity} discusses the implications and scope of this benchmark-informed setting.

%% file: sections/approach.tex
\section{Approach}
\label{sec:approach}

\begin{figure*}[t]
  \centering
  \includegraphics[width=0.9\textwidth]{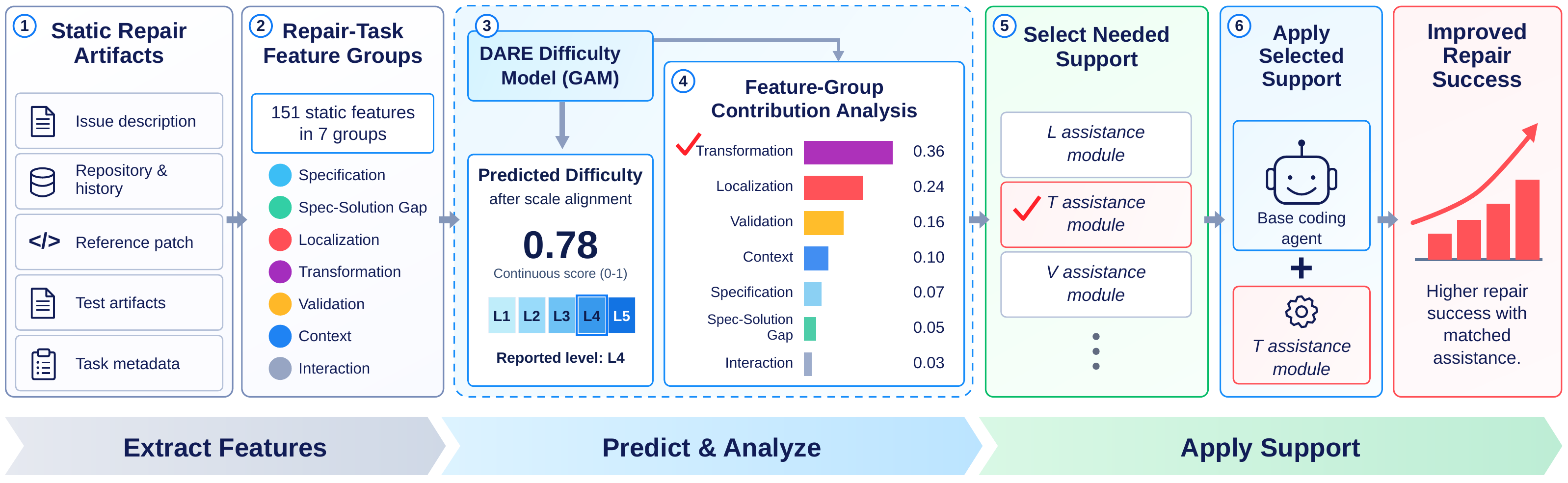}
  \caption{Overview of the \aname{} framework. Static repair artifacts yield task features organized into seven Feature Groups. The \aname{} difficulty model predicts repair difficulty, while its feature contributions are aggregated by Feature Group to guide the selection of task-specific assistance. 
  The displayed contribution values and selected module are illustrative.}
  \label{fig:dare_overview}
\end{figure*}

Figure~\ref{fig:dare_overview} shows the overview of \aname{}, a framework
that connects repair-task difficulty prediction, Feature Group contribution
analysis, and task-specific agent support. For each repair task, \aname{}
answers three questions in sequence: \emph{How difficult is this task?},
\emph{Which Feature Group is most salient in the model's difficulty
prediction?}, and \emph{What type of support could help the agent solve it?}

\aname{} first (1)~takes static repair artifacts, such as the issue
description, repository history, reference patch, and test artifacts, and
(2)~extracts features organized into seven Feature Groups: S, SG, L, T, V, C,
and INT (Section~\ref{sec:repair_task_metrics}). To answer the first question, (3)~the \aname{} difficulty model
predicts the task's difficulty score, also reported as levels L1--L5. To answer the
second, (4)~\aname{} aggregates the model's feature-level contributions by
Feature Group to identify the most salient group. To answer the third,
(5)~\aname{} selects the assistance module matching that group, and (6)~the
coding agent attempts the repair with this module. In the example,
Transformation (T) is the most salient group, so the agent receives the
T assistance module to improve its repair success. The following sections detail the key steps.



\subsection{Repair-Task Difficulty}
\label{sec:repair_task_difficulty}
To predict how difficult a repair task is, we first need a measurable notion of difficulty.

\noindent\textbf{Difficulty Definition.}
Response-based difficulty measures are established in educational assessments (proportion correct)~\cite{kim2024itemdifficulty}, human reliability research (error probability and task complexity)~\cite{podofillini2013complexity}, and software repair studies (coding-agent outcomes)~\cite{alhaque2026difficulty}. Following this approach, we define the repair difficulty of task $i$ as
\begin{equation}
  D_i=\frac{\text{Failed agents on task }i}{\text{Agents that evaluated task }i}.
  \label{eq:task_difficulty}
\end{equation}
Here, $D_i\in[0,1]$; larger values indicate higher agent failure rates. Tasks not evaluated by an agent are excluded rather than counted as failures. 
We use SWE-bench Verified to establish the reference difficulty for cross-population consistency analysis because it provides extensive and complete evaluation data: 125 of its 175 agent configurations evaluated all 500 tasks.

\noindent\textbf{Cross-Population Consistency.}
A common difficulty measure requires that tasks keep similar relative difficulty across agent populations. We split the 125 complete Verified configurations along four dimensions: provider (12 OpenAI vs.\ 9 Anthropic, same mini-SWE-agent scaffold), scaffold (41 mini-SWE-agent vs.\ 84 others), submission era (split at the median submission date), and performance level (split at the median overall solve rate). For each split, we compute each group's difficulty for all 500 tasks independently and measure the Spearman rank correlation between the two task-aligned vectors (Table~\ref{tab:cross_population_consistency}).

All four comparisons agree strongly (${\rho=.808\text{--}.921}$), even
between higher- and lower-performing agents (${\rho=.894}$). The lowest
value, for Provider, partly reflects its small groups, which yield coarser
estimates. As agents improve, absolute failure rates decline, much as exam
scores rise as students learn, but task ordering stays stable across
submission eras (${\rho=.921}$); scale alignment adjusts the absolute level
(Section~\ref{sec:scale_alignment}). This indicates a shared task-level
difficulty structure across agent populations, supporting multi-agent
failure rates as our difficulty measure.

\begin{table}[t]
  \centering
  \caption{Cross-population consistency of task difficulty on SWE-bench Verified.}
  \label{tab:cross_population_consistency}
  \begin{tabular}{lc}
    \toprule
    Population comparison & Spearman $\rho$ \\
    \midrule
    Provider & .808 \\
    Scaffold & .914 \\
    Submission era & .921 \\
    Performance level & .894 \\
    \bottomrule
  \end{tabular}
\end{table}

\noindent\textbf{Continuous Difficulty.}
We keep difficulty continuous throughout training and prediction, because
dichotomizing continuous variables loses information and statistical
power~\cite{altman2006dichotomising}. Training on multiple benchmarks, however, raises
two scale problems. First, benchmarks differ in their difficulty
distributions, so the same failure rate can be relatively easy in one
benchmark and relatively hard in another. We therefore train the model on
\emph{normalized} difficulty: within each benchmark, a task's failure rate
is converted to its relative rank and mapped to a standard normal value.
The model's predictions are then mapped back to the original $[0,1]$
failure-rate scale using the training benchmarks' difficulty distribution
(Section~\ref{sec:model}). Second, failure rates depend on which agents were evaluated, and training and target benchmarks involve different agent populations. \emph{Shared-agent scale alignment} uses two
anchor agents evaluated on all benchmarks to recalibrate the predictions to
the target benchmark's scale (Section~\ref{sec:scale_alignment}). The final
predictions $\hat{D}_i$ are evaluated against the remaining non-anchor
agents, so alignment and evaluation never use the same agents.

\begin{table}[t]
  \centering
  \caption{Fixed absolute difficulty levels.}
  \label{tab:absolute_difficulty_levels}
  \begin{tabular}{ll}
    \toprule
    Level & Difficulty interval \\
    \midrule
    L1 & $0\le \hat{D}_i<0.2$ \\
    L2 & $0.2\le \hat{D}_i<0.4$ \\
    L3 & $0.4\le \hat{D}_i<0.6$ \\
    L4 & $0.6\le \hat{D}_i<0.8$ \\
    L5 & $0.8\le \hat{D}_i\le1.0$ \\
    \bottomrule
  \end{tabular}
\end{table}

\noindent\textbf{Absolute Difficulty Levels.}
To check whether predicted difficulty agrees with observed failure rates
(RQ1) and to identify tasks predicted to be difficult (RQ2), we group the
final predictions into levels. Following the observation that four or five categories can limit information loss when grouping is necessary~\cite{altman2006dichotomising}, we divide
$[0,1]$ into five equal-width intervals of 0.2, labeled L1--L5
(Table~\ref{tab:absolute_difficulty_levels}), as in calibration
studies~\cite{guo2017calibration}. Unlike benchmark-relative groups such as
quintiles, fixed intervals denote the same failure-rate range in every
benchmark, allowing comparison on a common scale and assessment of absolute
agreement, not only ordering~\cite{vancalster2019calibration}. Levels are
used only for interpretation, never for training. Because failure rates
depend on the evaluated agents, a level describes difficulty for the target
benchmark's agent population rather than a population-independent task
property.

\subsection{Repair-Task Feature Groups}
\label{sec:repair_task_metrics}
Task difficulty indicates which tasks are difficult, but not why. Software repair requires understanding the problem, locating relevant code, making changes, and validating them~\cite{gazzola2019survey}. Prior work has examined bug-report information~\cite{bettenburg2008goodbugreport}, fault localization~\cite{zhou2012buglocalization}, code-change size and characteristics~\cite{kamei2013jitqa}, test-based validation~\cite{legoues2012genprog}, and links between static task characteristics and coding-agent performance~\cite{alhaque2026difficulty}.
We organize metrics from GitHub issue descriptions, repository snapshots and Git history, reference patches, and test artifacts into seven Repair-Task Feature Groups reflecting the repair process.

\noindent\textbf{Specification ($S$)} captures issue-description clues for understanding a problem and its requirements. Metrics cover issue length, structure, readability, error messages, stack traces, reproduction steps, and code identifiers. Prior work shows that reproduction steps and stack traces can aid bug resolution~\cite{bettenburg2008goodbugreport}.

\noindent\textbf{Specification--Solution Gap ($\mathit{SG}$)} measures how directly issue information relates to the changes required for a solution~\cite{tian2022quatrain}. Its metrics include textual similarity between the issue and reference patch or test artifacts, identifier overlap, solution identifier novelty, and specification coverage. Unlike $S$, which characterizes the issue itself, $\mathit{SG}$ explicitly captures the information gap between the problem specification and solution artifacts as a separate group.

\noindent\textbf{Localization ($L$)} characterizes clues for identifying the files and code locations requiring modification. Its metrics include file paths, modules, and symbols mentioned in the issue, matches between stack traces and modified locations, and modified-file ranks in BM25- and TF-IDF-based retrieval results. Prior research has used information retrieval to locate relevant source files from bug reports~\cite{zhou2012buglocalization}. Building on this approach, $L$ measures the correspondence between clues provided in the issue and the actual modification locations.

\noindent\textbf{Transformation ($T$)} characterizes the size and structural complexity of code changes required to resolve a repair task. Its metrics capture the scope and distribution of reference-patch changes, modified functions and classes, and changes to function signatures, conditionals, loops, and imports. Prior research has linked code change characteristics, including size and diffusion, to software defect risk~\cite{kamei2013jitqa}. We use these characteristics to represent the extent and structure of code transformations involved in repair.

\noindent\textbf{Validation ($V$)} characterizes the requirements for verifying that a patch resolves an issue without breaking existing functionality. Metrics include \texttt{FAIL\_TO\_PASS} and \texttt{PASS\_TO\_PASS} test composition, test-patch size and distribution, assertions, and fixtures, capturing validation scope and structural complexity. Test suites commonly serve as behavioral oracles in APR~\cite{legoues2012genprog}.
Thus, $V$ measures the scope and structural complexity of a task’s validation artifacts.

\noindent\textbf{Context ($C$)} characterizes the structure and development history of the repository containing the code to be modified, beyond the size of an individual patch. Metrics cover file-path structure, dependency-graph fan-in and fan-out, Git change frequency and recency, and task metadata, reflecting connections to surrounding components and past changes~\cite{zimmermann2008dependency,zimmermann2004rose}.

\noindent\textbf{Interaction ($\mathit{INT}$)} captures combinations of static characteristics across groups, such as specification--solution gap with patch complexity, localization features with repository size, and change history with dependency-graph centrality. These combinations can reflect characteristics missed by individual features; $\mathit{INT}$ denotes static-feature interactions, not interactions between an agent and its tools or environment.

Our contribution is to organize previously studied repair-related factors into a unified measurement framework comprising seven Feature Groups. Each feature group reflects a distinct aspect of the software repair process. We aggregate feature-level contributions within these groups to characterize the model's difficulty predictions and identify each task's most salient Feature Group (Section~\ref{sec:dare}). Feature extraction and selection are described in Section~\ref{sec:features}.

\subsection{Difficulty Model and Feature Group Contributions}
\label{sec:dare}

The \aname{} difficulty model employs GAM-style additive regression to capture nonlinear relationships between features and difficulty while retaining interpretable feature-level contributions (Section~\ref{sec:model}). For task $i$ with feature vector $\mathbf{x}_i=(x_{i1},\ldots,x_{ip})$, the additive prediction score is
\begin{equation}
  s_i=\beta_0+\sum_{j=1}^{p}f_j(x_{ij}).
  \label{eq:dare_score}
\end{equation}
Here, $i$ indexes tasks, $j$ indexes features, and $p$ is the total number of features. The term $x_{ij}$ denotes the value of feature $j$ for task $i$, $\beta_0$ is the intercept, and $f_j(x_{ij})$ is the learned contribution of feature $j$ to the prediction score. Positive contributions increase the prediction score, whereas negative contributions decrease it. The final difficulty estimate $\hat{D}_i$ is obtained by transforming this score to the $[0,1]$ failure-rate scale and applying shared-agent scale alignment (Sections~\ref{sec:model} and~\ref{sec:scale_alignment}). Feature contributions apply to the score before these transformations, not directly to $\hat{D}_i$.

We aggregate the difficulty model's feature-level contributions within each Feature Group. Let $\mathcal{J}_g$ be the features in group $g$. Its contribution to the prediction score for task $i$ is
\begin{equation}
  C_{i,g}=\sum_{j\in\mathcal{J}_g}f_j(x_{ij}).
  \label{eq:group_contribution}
\end{equation}
Each task therefore has both a predicted difficulty $\hat{D}_i$ and a Feature Group contribution profile:
\begin{equation}
  \mathbf{C}_i=(C_{i,S},C_{i,\mathit{SG}},C_{i,L},C_{i,T},C_{i,V},C_{i,C},C_{i,\mathit{INT}}).
  \label{eq:group_profile}
\end{equation}
For example, large contributions from $L$ and $V$ indicate that features related to locating the code to modify and validating the resulting patch contribute strongly to the prediction score.
For each group, we subtract its average contribution over the training tasks, so that a positive centered contribution indicates that the group raises a task's predicted difficulty more than it does for a typical training task.
These centered values are model-based attributions, not causal evidence of repair failure.

Even tasks with similar predicted difficulty may have different Feature Group contribution profiles. We define a task's \emph{salient Feature Group} as the group with the largest positive centered contribution; a task is labeled \emph{None} if no group has a positive centered contribution.

\subsection{Feature-Group-Guided Repair Support}
\label{sec:support}

Because tasks with similar predicted difficulty can have different salient groups, \aname{} uses the salient group, rather than the difficulty score alone, to select task-specific support. Each Feature Group can be paired with an assistance module that targets the repair activity the group characterizes; the module gives the agent additional analysis and feedback without changing the underlying model. For each task, \aname{} attaches the module for its salient group. Tasks labeled \emph{None}, or whose salient group has no available module, are attempted without assistance. Modules use only information available to the agent, never reference patches or hidden tests.

For example, when $T$ is the salient group (Fig.~\ref{fig:dare_overview}), the task is expected to require large or structurally complex changes. The T-Module therefore helps the agent plan dependency-aware follow-up changes; after a function signature is changed, it points to call sites that may also need updating. Section~\ref{sec:repair_support} describes the modules implemented in our evaluation.

Thus, \aname{}'s contribution lies not in the additive model itself, but in integrating difficulty prediction, Feature Group analysis, and task-specific support within a single framework.

%% file: sections/experimental_setup.tex
\section{Experimental Setup} \label{sec:setup}

\subsection{Feature Extraction and Selection}
\label{sec:features}
We extract static features to characterize repair tasks and organize them
into the seven Feature Groups (Section~\ref{sec:repair_task_metrics}). They are
computed from issue descriptions, repository snapshots and Git history,
reference patches, test artifacts, and task metadata. Repository features
are extracted from the state at each task's \texttt{base\_commit}, using
only Git history reachable from that commit. Because our features include
information from reference patches and test artifacts, we target difficulty
prediction during benchmark construction and evaluation, before large-scale
agent evaluation. Accordingly, feature extraction uses no agent
trajectories, repair outcomes, or observed difficulty, which also prevents
label leakage.

We start with 268 predefined candidate metrics and apply four feature selection steps to reduce redundancy and multicollinearity. First, we remove features with zero variance. Next, we remove highly correlated features within each Feature Group using a Spearman correlation threshold of $|\rho|>0.90$. The 0.90 threshold, previously used for correlation-based feature filtering~\cite{kuhn2008caret}, removes only very strong pairwise associations. We then remove near-exact linear dependencies ($R^2>1-10^{-10}$), corresponding to an implied variance inflation factor (VIF) greater than $10^{10}$, to reduce the risk of numerical instability before VIF-based filtering. Here, $10^{-10}$ serves as a numerical tolerance rather than a statistical significance threshold. Finally, we iteratively remove features with a VIF greater than 10, following a commonly used heuristic for multicollinearity screening~\cite{dormann2013collinearity,obrien2007vif}. This process produces a final set of 151 features.

We extract features consistently across all four benchmark partitions. The final 151 features are selected using only SWE-bench Verified feature values, without difficulty labels or agent outcomes, so no label information leaks even when Verified is held out, and are applied to all four partitions. The selected features comprise 24 $S$, 17 $\mathit{SG}$, 15 $L$, 46 $T$, 29 $V$, 15 $C$, and 5 $\mathit{INT}$ features. During model training and cross-validation, missing-value handling and feature transformations are fitted using only the training data, as described in Section~\ref{sec:model}. The complete feature list, including individual definitions, group assignments, and extraction procedures, is provided in the accompanying replication artifact~\cite{dare_replication}.

\subsection{Difficulty Model Training}
\label{sec:model}

We implement the \aname{} difficulty model as a regularized Generalized Additive Model (GAM), which provides both difficulty predictions and feature-level contributions (Section~\ref{sec:dare}). GAM is a widely established interpretable model that captures nonlinear relationships while allowing each feature's contribution to the prediction to be computed directly~\cite{hastie1986gam,lou2012intelligible}. Although other additive models, such as Explainable Boosting Machines (EBMs) and Neural Additive Models (NAMs), offer similar capabilities~\cite{nori2019interpretml,agarwal2021nam}, we choose GAM for its relatively simple structure, controlled model complexity, and clear feature-level contributions. Specifically, we use cubic splines to capture nonlinear relationships and ridge regularization to control model complexity~\cite{hoerl1970ridge}. The additive structure also allows feature contributions to be aggregated within each feature group, enabling the same difficulty model to support both prediction and feature-group analysis.

The model uses 151 features across the seven Feature Groups. To account for differing difficulty distributions across benchmarks, we convert observed difficulty within each source partition to ranks and map them to standard normal quantiles. This places difficulty values from different benchmarks on a common scale during training.

We select model hyperparameters using repository-grouped 5-fold cross-validation repeated 10 times, following the use of repeated five-fold validation in prior software prediction studies~\cite{gao2015aggregating,ali2021validation}. The five-fold setting balances training-set size and validation coverage, while repetition reduces sensitivity to a particular fold partition. Grouping by repository prevents tasks from the same repository from appearing in both training and validation sets within a fold, thereby accounting for dependencies among tasks~\cite{roberts2017crossvalidation}. Within each cross-validation split, feature preprocessing and difficulty normalization are fitted using only the training data and then applied to the validation data. After hyperparameter selection, the model is retrained on the complete source training data. Its prediction scores are then mapped back to the $[0,1]$ failure-rate scale using the observed difficulty distribution of the source training data. Shared-agent scale alignment is then applied to obtain the final difficulty predictions. The alignment procedure is described in Section~\ref{sec:scale_alignment}.

\subsection{Shared-Agent Scale Alignment}
\label{sec:scale_alignment}

As established in Section~\ref{sec:repair_task_difficulty}, task difficulty can vary depending on the population of coding agents being evaluated. To account for these differences, we apply Shared-Agent Scale Alignment to map predictions from the difficulty model trained on source benchmarks to the failure-rate scale of the target benchmark. This approach builds on logistic recalibration, which adjusts the intercept and slope of a prediction model when applying it to a new population~\cite{steyerberg2004validation,vancalster2019calibration}.

We use two SWE-agent-based configurations, Claude 3.7 Sonnet and GPT-4o, as anchor agents because their results are available across all four benchmarks. We first generate initial difficulty predictions, $\tilde{D}_i$, for target tasks using the model trained on the source data, as described in Section~\ref{sec:model}. We then fit a logistic regression using the observed success (1) or failure (0) of each anchor agent, denoted by $Y_{ia}$:
\begin{equation}
\begin{array}{@{}r@{}l@{}}
\mathrm{logit}\,P(Y_{ia}=1)&=\alpha+\beta z_i+s_a\delta,\\
z_i&=\mathrm{logit}(1-\tilde{D}_i).
\end{array}
\label{eq:shared_agent_alignment}
\end{equation}

Here, $\alpha$ adjusts the overall success-rate level, $\beta$ controls the spread of predictions, and $\delta$ accounts for performance differences between the two anchor agents. We set $s_a=+1$ for Claude 3.7 Sonnet and $s_a=-1$ for GPT-4o. For each held-out benchmark, the three parameters are estimated by maximum likelihood using only the observed anchor-agent outcomes. We constrain $\beta=\exp(\theta)>0$ and clip logit inputs to $[10^{-6},1-10^{-6}]$ for numerical stability. Missing agent outcomes are excluded.

The final difficulty prediction is obtained from the fitted relationship without the agent-specific effect:
\begin{equation}
\hat{D}_i=1-\sigma(\hat{\alpha}+\hat{\beta}z_i),
\label{eq:aligned_difficulty}
\end{equation}
where $\sigma$ denotes the sigmoid function. This monotonic transformation adjusts the absolute failure-rate scale without reversing the predicted ordering of tasks. Accordingly, the feature-group contributions defined in Section~\ref{sec:dare} are interpreted using the additive prediction scores before alignment.

To separate alignment from evaluation, we do not use the aggregate difficulty labels or non-anchor agent outcomes of the held-out benchmark for model training or alignment. Only the two anchor agents' observed outcomes are used for alignment, while final performance is evaluated against the failure rates of the remaining non-anchor agents. This allows us to evaluate difficulty predictions against agents not used for alignment.

\subsection{Evaluation Metrics}
\label{sec:evaluation_metrics}

We evaluate the final aligned predictions from the \aname{} difficulty model using three complementary metrics: Spearman rank correlation ($\rho$), pairwise accuracy (pAcc), and mean absolute error (MAE).

\noindent\textbf{Spearman Rank Correlation ($\rho$).}
Spearman $\rho$ measures the rank agreement between predicted and observed task difficulty by capturing monotonic relationships~\cite{spearman1904}. It has also been used to evaluate difficulty prediction across benchmarks~\cite{krsteski2026predicting}.

\noindent\textbf{Pairwise Accuracy (pAcc).}
pAcc measures the proportion of task pairs correctly ordered by predicted difficulty. Let $D_i$ and $\hat{D}_i$ denote the observed and predicted difficulty of task $i$, respectively. For comparable pairs $\mathcal{P}=\{(i,j)\mid i<j,\,D_i\neq D_j\}$, pAcc is defined as
\begin{equation}
\mathrm{pAcc}=\frac{1}{|\mathcal{P}|}\sum_{(i,j)\in\mathcal{P}}
\mathbf{1}\!\left[(\hat{D}_i-\hat{D}_j)(D_i-D_j)>0\right].
\label{eq:pacc}
\end{equation}
Pairs tied in observed difficulty are excluded; predicted ties on otherwise comparable pairs are counted as incorrect. Krsteski and Meyer~\cite{krsteski2026predicting} also used within-benchmark pAcc alongside Spearman correlation as a complementary metric. We therefore report both to assess overall ranking agreement and pairwise difficulty discrimination.

\noindent\textbf{Mean Absolute Error (MAE).}
MAE averages the absolute differences between predicted and observed task difficulty, indicating how closely predictions match observations~\cite{willmott2005mae}:
\begin{equation}
\mathrm{MAE}=\frac{1}{N}\sum_{i=1}^{N}|D_i-\hat{D}_i|,
\label{eq:mae}
\end{equation}
where $N$ is the number of evaluation tasks and $\hat{D}_i$ is the final aligned prediction. Because difficulty is measured as a failure rate on $[0,1]$, an MAE of $0.15$ corresponds to an average error of 15 percentage points.

Together, the three metrics provide complementary assessments: Spearman $\rho$ captures overall difficulty ranking agreement, pAcc measures pairwise discrimination, and MAE quantifies absolute prediction error. We thus evaluate both the relative ordering and the absolute difficulty of repair tasks.

For RQ1, we obtain 95\% confidence intervals (CIs) by task-level bootstrap resampling, using 2,000 resamples for Spearman's $\rho$ and 300 for the costlier pairwise pAcc.

\subsection{Benchmarks}
\label{sec:benchmarks}
We use SWE-bench and SWE-bench-Live, two benchmarks built from real-world GitHub issues. Both provide repository-level software repair tasks with reference patches and executable tests for evaluating the correctness of proposed fixes~\cite{jimenez2024swebench,zhang2025swebenchgoeslive}.

\noindent\textbf{SWE-bench.}
SWE-bench evaluates coding agents on software repair tasks derived from real GitHub issues and their associated pull requests~\cite{jimenez2024swebench}. The Full test set contains 2,294 tasks collected from 12 Python repositories. For each task, an agent must generate a patch using the issue description and the repository state before the fix, and the patch is evaluated using executable tests~\cite{jimenez2024swebench}. 

To support different evaluation needs, SWE-bench provides two curated subsets of the Full test set: Lite for lower-cost evaluation and Verified for more reliable task selection. SWE-bench Lite is a 300-task subset designed to reduce evaluation costs while covering a range of task difficulties, with an emphasis on relatively self-contained bug fixes~\cite{swebench2024lite}. SWE-bench Verified is a 500-task subset reviewed by professional developers to reduce evaluation issues caused by ambiguous problem descriptions or inadequate tests~\cite{chowdhury2024verified}.

Because Verified and Lite are both subsets of Full, using these datasets directly could introduce overlapping tasks between training and evaluation. To prevent this, we partition SWE-bench by \texttt{instance\_id} into Verified (500 tasks), Lite $\setminus$ Verified (207 tasks), and Full $\setminus$ (Verified $\cup$ Lite) (1,587 tasks), as shown in Table~\ref{tab:benchmark_partitions}. These mutually exclusive partitions cover all 2,294 tasks in Full.

\noindent\textbf{SWE-bench-Live.}
To evaluate \aname{} on more recently collected real-world repair tasks beyond those in the original SWE-bench, we use SWE-bench-Live Lite, a fixed split of 300 Python tasks collected from GitHub~\cite{zhang2025swebenchgoeslive,swebenchlive2025repository}. This split constitutes the fourth evaluation partition in Table~\ref{tab:benchmark_partitions}. Unlike SWE-bench-Live's continuously updated Full split, the fixed Lite split enables reproducible evaluations while keeping repeated agent executions in RQ3 manageable. We use this split to evaluate difficulty prediction in RQ1 and to assess whether support informed by the RQ2 feature-group analysis improves coding-agent repair success in RQ3.

\noindent\textbf{Agent Outcomes and Evaluation Partitions.}
We construct task difficulty from publicly available agent evaluation results, using official SWE-bench evaluation records and SWE-bench-Live submission records~\cite{swebench2024experiments,swebenchlive2025repository}. SWE-bench Verified contains results from 175 agent configurations, of which 125 have outcomes for all 500 tasks. We use these 125 complete configurations to construct the reference difficulty for Verified. For the remaining partitions, we use the observed outcomes from 84 configurations for Lite $\setminus$ Verified, 24 for Full $\setminus$ (Verified $\cup$ Lite), and 20 for SWE-bench-Live Lite. Missing task-agent outcomes are excluded rather than treated as failures.

For RQ1, we use leave-one-dataset-out (LODO) evaluation, training the \aname{} difficulty model on three partitions and evaluating its aligned predictions on the fourth. Unlike random task splits, LODO evaluates generalization to held-out task sets, following prior cross-dataset and cross-project validation studies~\cite{laroca2022crossdataset,hosseini2019crossproject}. Task overlap between training and evaluation partitions is removed, although some repositories are shared across partitions. The separation of scale alignment and evaluation is described in Section~\ref{sec:scale_alignment}.

\begin{table}[!t]
\centering
\caption{Benchmarks and LODO evaluation partitions.}
\label{tab:benchmark_partitions}
\footnotesize
\setlength{\tabcolsep}{2pt}
\begin{tabular*}{\columnwidth}{@{}p{0.43\columnwidth}@{\extracolsep{\fill}}rrr@{}}
\toprule
Benchmark / partition & \#Test & \#Train & \#Agents \\
\midrule
SWE-bench Verified & 500 & 2,094 & 125 \\
Lite $\setminus$ Verified & 207 & 2,387 & 84 \\
Full $\setminus$ (Verified $\cup$ Lite) & 1,587 & 1,007 & 24 \\
SWE-bench-Live Lite & 300 & 2,294 & 20 \\
\bottomrule
\end{tabular*}
\par\vspace{2pt}
\noindent\parbox{\columnwidth}{\footnotesize\emph{Note.} \#Test is the held-out task count; \#Train is the count in the other three partitions. \#Agents counts public configurations used to construct difficulty, not benchmark-defined totals. RQ1 evaluation excludes alignment anchor agents; RQ3 intervention procedures are described in Section~\ref{sec:rq3}.}
\end{table}

%% file: sections/results.tex
\section{Results}\label{sec:results}
\input{sections/rq1}

\input{sections/rq2}

\input{sections/rq3}

%% file: sections/rq1.tex
\subsection{RQ1: How Accurately Does \aname{} Predict Repair Difficulty?}
\label{sec:rq1}

\subsubsection{Prediction Accuracy on Held-Out Benchmarks}

Across four LODO holdouts, the \aname{} difficulty model consistently predicted relative task-difficulty rankings with moderate positive rank correlations and above-chance pairwise discrimination (Table~\ref{tab:rq1_results}). We assess performance using Spearman's $\rho$, pairwise accuracy (pAcc), and mean absolute error (MAE); 95\% confidence intervals (CIs) are computed as described in Section~\ref{sec:evaluation_metrics}. Predictions are aligned using two shared anchor agents and evaluated against non-anchor agents not used for alignment (Section~\ref{sec:scale_alignment}).

\begin{table}[!t]
\centering
\caption{Difficulty prediction performance under cross-partition holdout evaluation.}
\label{tab:rq1_results}
\scriptsize
\setlength{\tabcolsep}{2.2pt}
\renewcommand{\arraystretch}{1.13}
\begin{tabular*}{\columnwidth}{@{}l@{\extracolsep{\fill}}ccc@{}}
\toprule
Held-out partition &
$\rho$ [95\% CI] & pAcc [95\% CI] & MAE \\
\midrule
SWE-bench Verified & .621 [.563,.674] & .723 [.698,.748] & .254 \\
Lite $\setminus$ Verified & .541 [.434,.635] & .694 [.658,.737] & .197 \\
Full $\setminus$ (Verified$\cup$Lite) & .566 [.530,.601] & .744 [.725,.759] & .123 \\
SWE-bench-Live Lite & .476 [.376,.564] & .675 [.639,.708] & .149 \\
\bottomrule
\end{tabular*}
\end{table}

\begin{figure}[!t]
\centering
\resizebox{\columnwidth}{!}{%
\begin{tikzpicture}[x=0.1545pt,y=-0.1545pt,inner sep=0pt,outer sep=0pt]
  \path[use as bounding box] (0,0) rectangle (2880,2200);
  \definecolor{rqBox}{RGB}{207,224,243}
  \definecolor{rqBlue}{RGB}{43,93,138}
  \definecolor{rqRed}{RGB}{192,57,43}
  \definecolor{rqGrid}{RGB}{221,221,221}
  \newcommand{\drawrqbox}[6]{%
    \draw[line width=.67pt,color=rqBlue] (#1,#2) -- (#1,#3);%
    \draw[line width=.67pt,color=rqBlue] ({#1-38},#2) -- ({#1+38},#2);%
    \draw[line width=.67pt,color=rqBlue] (#1,#5) -- (#1,#6);%
    \draw[line width=.67pt,color=rqBlue] ({#1-38},#6) -- ({#1+38},#6);%
    \filldraw[fill=rqBox,draw=rqBlue,line width=.67pt]
      ({#1-74},#3) rectangle ({#1+74},#5);%
    \draw[color=rqRed,line width=1.15pt] ({#1-74},#4) -- ({#1+74},#4);%
  }
  \draw[color=rqGrid,line width=.37pt] (199,330) -- (1487,330);
  \draw[color=rqGrid,line width=.37pt] (199,487) -- (1487,487);
  \draw[color=rqGrid,line width=.37pt] (199,643) -- (1487,643);
  \draw[color=rqGrid,line width=.37pt] (199,800) -- (1487,800);
  \draw[color=black,line width=.65pt] (199,173) rectangle (1487,957);
  \node[font=\footnotesize,anchor=center] at (843,77) {SWE-bench Verified};
  \draw[black,line width=.55pt] (347,957) -- (347,968);
  \node[font=\scriptsize,anchor=north] at (347,1003) {L1};
  \node[font=\scriptsize,anchor=center] at (347,132) {n=11};
  \draw[black,line width=.55pt] (595,957) -- (595,968);
  \node[font=\scriptsize,anchor=north] at (595,1003) {L2};
  \node[font=\scriptsize,anchor=center] at (595,132) {n=110};
  \draw[black,line width=.55pt] (843,957) -- (843,968);
  \node[font=\scriptsize,anchor=north] at (843,1003) {L3};
  \node[font=\scriptsize,anchor=center] at (843,132) {n=156};
  \draw[black,line width=.55pt] (1090,957) -- (1090,968);
  \node[font=\scriptsize,anchor=north] at (1090,1003) {L4};
  \node[font=\scriptsize,anchor=center] at (1090,132) {n=121};
  \draw[black,line width=.55pt] (1338,957) -- (1338,968);
  \node[font=\scriptsize,anchor=north] at (1338,1003) {L5};
  \node[font=\scriptsize,anchor=center] at (1338,132) {n=102};
  \draw[black,line width=.55pt] (189,957) -- (199,957);
  \node[font=\scriptsize,anchor=east] at (171,957) {0.0};
  \draw[black,line width=.55pt] (189,800) -- (199,800);
  \node[font=\scriptsize,anchor=east] at (171,800) {0.2};
  \draw[black,line width=.55pt] (189,643) -- (199,643);
  \node[font=\scriptsize,anchor=east] at (171,643) {0.4};
  \draw[black,line width=.55pt] (189,487) -- (199,487);
  \node[font=\scriptsize,anchor=east] at (171,487) {0.6};
  \draw[black,line width=.55pt] (189,330) -- (199,330);
  \node[font=\scriptsize,anchor=east] at (171,330) {0.8};
  \draw[black,line width=.55pt] (189,173) -- (199,173);
  \node[font=\scriptsize,anchor=east] at (171,173) {1.0};
  \node[rotate=90,font=\footnotesize] at (68,565) {Observed non-anchor difficulty};
  \drawrqbox{347}{898}{905}{913}{920}{939}
  \drawrqbox{595}{519}{744}{846}{896}{939}
  \drawrqbox{843}{176}{540}{770}{860}{933}
  \drawrqbox{1090}{176}{333}{584}{740}{920}
  \drawrqbox{1338}{176}{183}{305}{598}{888}
  \draw[color=rqGrid,line width=.37pt] (1547,330) -- (2835,330);
  \draw[color=rqGrid,line width=.37pt] (1547,487) -- (2835,487);
  \draw[color=rqGrid,line width=.37pt] (1547,643) -- (2835,643);
  \draw[color=rqGrid,line width=.37pt] (1547,800) -- (2835,800);
  \draw[color=black,line width=.65pt] (1547,173) rectangle (2835,957);
  \node[font=\footnotesize,anchor=center] at (2190,77) {Lite $\setminus$ Verified};
  \draw[black,line width=.55pt] (1689,957) -- (1689,968);
  \node[font=\scriptsize,anchor=north] at (1689,1003) {L1};
  \node[font=\scriptsize,text=gray,anchor=center] at (1689,132) {n=0};
  \draw[black,line width=.55pt] (1937,957) -- (1937,968);
  \node[font=\scriptsize,anchor=north] at (1937,1003) {L2};
  \node[font=\scriptsize,anchor=center] at (1937,132) {n=8};
  \draw[black,line width=.55pt] (2185,957) -- (2185,968);
  \node[font=\scriptsize,anchor=north] at (2185,1003) {L3};
  \node[font=\scriptsize,anchor=center] at (2185,132) {n=28};
  \draw[black,line width=.55pt] (2432,957) -- (2432,968);
  \node[font=\scriptsize,anchor=north] at (2432,1003) {L4};
  \node[font=\scriptsize,anchor=center] at (2432,132) {n=70};
  \draw[black,line width=.55pt] (2680,957) -- (2680,968);
  \node[font=\scriptsize,anchor=north] at (2680,1003) {L5};
  \node[font=\scriptsize,anchor=center] at (2680,132) {n=101};
  \drawrqbox{1937}{219}{307}{641}{811}{871}
  \drawrqbox{2185}{181}{434}{575}{763}{861}
  \drawrqbox{2432}{176}{202}{351}{645}{880}
  \drawrqbox{2680}{176}{176}{202}{285}{450}
  \begin{scope}[shift={(0,34)}]
  \draw[color=rqGrid,line width=.37pt] (199,1358) -- (1487,1358);
  \draw[color=rqGrid,line width=.37pt] (199,1515) -- (1487,1515);
  \draw[color=rqGrid,line width=.37pt] (199,1671) -- (1487,1671);
  \draw[color=rqGrid,line width=.37pt] (199,1828) -- (1487,1828);
  \draw[color=black,line width=.65pt] (199,1201) rectangle (1487,1985);
  \node[font=\footnotesize,anchor=center] at (843,1110) {Full $\setminus$ (Verified $\cup$ Lite)};
  \draw[black,line width=.55pt] (347,1985) -- (347,1996);
  \node[font=\scriptsize,anchor=north] at (347,2032) {L1};
  \node[font=\scriptsize,text=gray,anchor=center] at (347,1166) {n=0};
  \draw[black,line width=.55pt] (595,1985) -- (595,1996);
  \node[font=\scriptsize,anchor=north] at (595,2032) {L2};
  \node[font=\scriptsize,anchor=center] at (595,1166) {n=17};
  \draw[black,line width=.55pt] (843,1985) -- (843,1996);
  \node[font=\scriptsize,anchor=north] at (843,2032) {L3};
  \node[font=\scriptsize,anchor=center] at (843,1166) {n=103};
  \draw[black,line width=.55pt] (1090,1985) -- (1090,1996);
  \node[font=\scriptsize,anchor=north] at (1090,2032) {L4};
  \node[font=\scriptsize,anchor=center] at (1090,1166) {n=287};
  \draw[black,line width=.55pt] (1338,1985) -- (1338,1996);
  \node[font=\scriptsize,anchor=north] at (1338,2032) {L5};
  \node[font=\scriptsize,anchor=center] at (1338,1166) {n=1180};
  \draw[black,line width=.55pt] (189,1985) -- (199,1985);
  \node[font=\scriptsize,anchor=east] at (171,1985) {0.0};
  \draw[black,line width=.55pt] (189,1828) -- (199,1828);
  \node[font=\scriptsize,anchor=east] at (171,1828) {0.2};
  \draw[black,line width=.55pt] (189,1671) -- (199,1671);
  \node[font=\scriptsize,anchor=east] at (171,1671) {0.4};
  \draw[black,line width=.55pt] (189,1515) -- (199,1515);
  \node[font=\scriptsize,anchor=east] at (171,1515) {0.6};
  \draw[black,line width=.55pt] (189,1358) -- (199,1358);
  \node[font=\scriptsize,anchor=east] at (171,1358) {0.8};
  \draw[black,line width=.55pt] (189,1201) -- (199,1201);
  \node[font=\scriptsize,anchor=east] at (171,1201) {1.0};
  \node[rotate=90,font=\footnotesize] at (68,1594) {Observed non-anchor difficulty};
  \drawrqbox{595}{1378}{1595}{1700}{1769}{1844}
  \drawrqbox{843}{1204}{1312}{1537}{1662}{1837}
  \drawrqbox{1090}{1204}{1238}{1379}{1555}{1879}
  \drawrqbox{1338}{1198}{1204}{1198}{1274}{1389}
  \draw[color=rqGrid,line width=.37pt] (1547,1358) -- (2835,1358);
  \draw[color=rqGrid,line width=.37pt] (1547,1515) -- (2835,1515);
  \draw[color=rqGrid,line width=.37pt] (1547,1671) -- (2835,1671);
  \draw[color=rqGrid,line width=.37pt] (1547,1828) -- (2835,1828);
  \draw[color=black,line width=.65pt] (1547,1201) rectangle (2835,1985);
  \node[font=\footnotesize,anchor=center] at (2190,1110) {SWE-bench-Live Lite};
  \draw[black,line width=.55pt] (1689,1985) -- (1689,1996);
  \node[font=\scriptsize,anchor=north] at (1689,2032) {L1};
  \node[font=\scriptsize,text=gray,anchor=center] at (1689,1166) {n=0};
  \draw[black,line width=.55pt] (1937,1985) -- (1937,1996);
  \node[font=\scriptsize,anchor=north] at (1937,2032) {L2};
  \node[font=\scriptsize,anchor=center] at (1937,1166) {n=1};
  \draw[black,line width=.55pt] (2185,1985) -- (2185,1996);
  \node[font=\scriptsize,anchor=north] at (2185,2032) {L3};
  \node[font=\scriptsize,anchor=center] at (2185,1166) {n=13};
  \draw[black,line width=.55pt] (2432,1985) -- (2432,1996);
  \node[font=\scriptsize,anchor=north] at (2432,2032) {L4};
  \node[font=\scriptsize,anchor=center] at (2432,1166) {n=72};
  \draw[black,line width=.55pt] (2680,1985) -- (2680,1996);
  \node[font=\scriptsize,anchor=north] at (2680,2032) {L5};
  \node[font=\scriptsize,anchor=center] at (2680,1166) {n=214};
  \drawrqbox{2185}{1331}{1421}{1570}{1799}{1942}
  \drawrqbox{2432}{1204}{1295}{1419}{1619}{1939}
  \drawrqbox{2680}{1204}{1249}{1293}{1373}{1549}
  \draw[color=rqRed,line width=1.15pt] (1863,1982) -- (2011,1982);
  \node[font=\footnotesize,anchor=center] at (843,2110) {Predicted absolute difficulty level};
  \node[font=\footnotesize,anchor=center] at (2190,2110) {Predicted absolute difficulty level};
  \end{scope}
\end{tikzpicture}
}
\caption{Distribution of observed non-anchor failure rates across five fixed predicted absolute-difficulty levels (L1--L5), assigned using final aligned predictions. Boxes indicate the interquartile range (IQR), center lines indicate medians, whiskers extend to the most extreme observations within $1.5\times\mathrm{IQR}$, and $n$ above each level indicates the number of tasks. Outlier markers are omitted for readability.}
\label{fig:rq1_boxplot}
\end{figure}
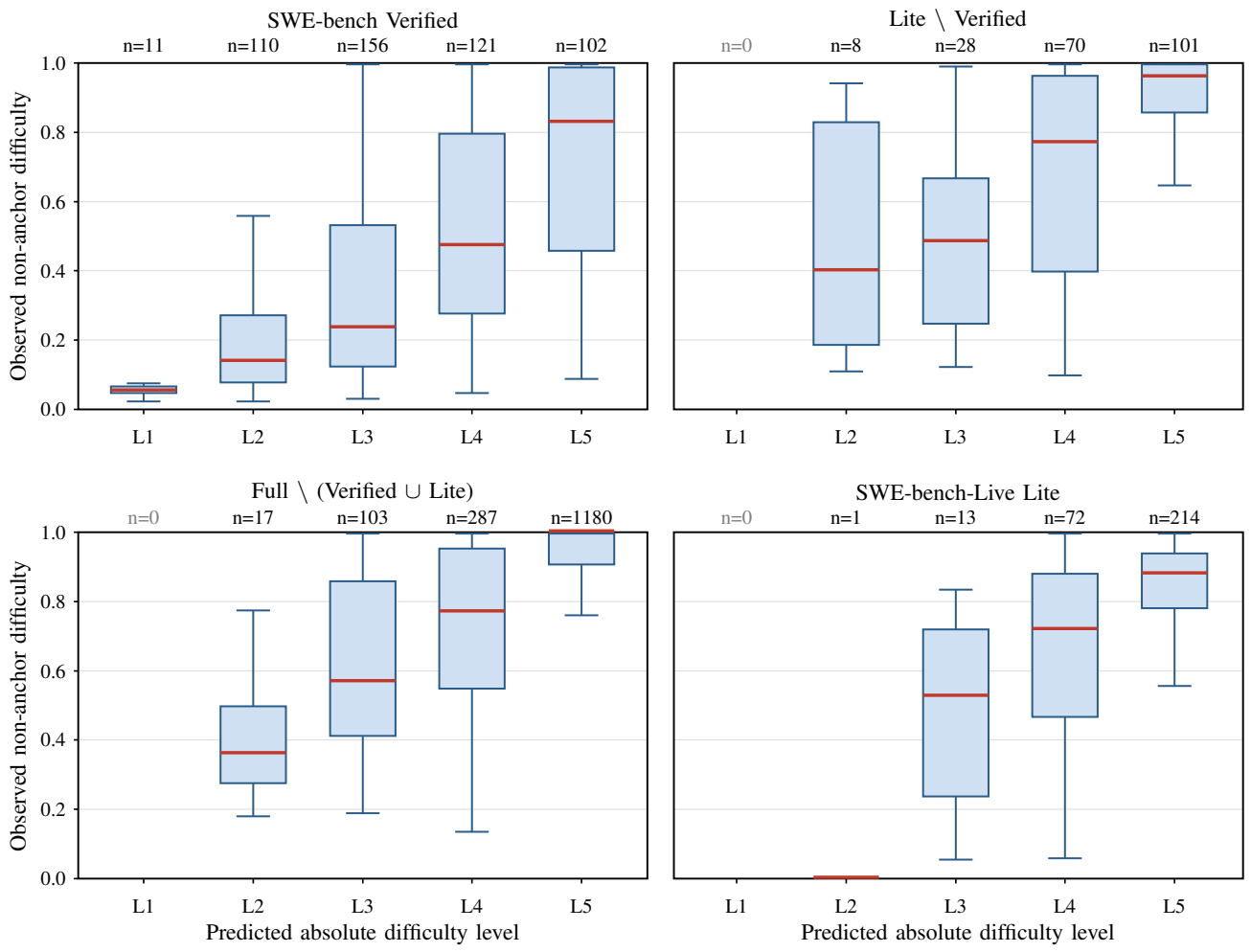

Spearman's $\rho$, pAcc, and MAE capture complementary aspects of prediction quality (Section~\ref{sec:evaluation_metrics}): task-difficulty rank agreement, pairwise discrimination, and absolute failure-rate error, respectively. Across four held-out partitions, $\rho$ ranged from .476 to .621, within the \emph{moderate} correlation range (.40--.69) of Schober et al.~\cite{schober2018correlation}, although this classification depends on context. All 95\% CIs lay above zero, indicating positive rank correlation rather than no association. This consistency across held-out partitions indicates that the difficulty model captures meaningful, though moderate, task-difficulty ordering.
pAcc=.675--.744 means that the model correctly identified the harder task in about 68--74\% of pairs with different observed difficulties. All 95\%
CIs were above .500, the chance level, supporting above-chance
discrimination in every partition. MAE=.123--.254 corresponds to mean absolute failure-rate errors of 12.3--25.4 percentage points.

For comparison, Krsteski and Meyer~\cite{krsteski2026predicting} reported mean $\rho$=.225 across 17 held-out agentic benchmarks; their best estimator ablation achieved $\rho$=.295 and pAcc=.602. Our model achieved mean $\rho$=.551 and pAcc=.709 across four partitions. However, different benchmark compositions and difficulty definitions preclude direct comparison. Verified had the highest $\rho$ (.621) but also the highest MAE (.254), showing that stronger rank agreement does not guarantee low absolute prediction error~\cite{vancalster2019calibration,steyerberg2010assessing,willmott2005mae}.

\subsubsection{Predicted Absolute Difficulty Levels Correspond to Observed Difficulty}

Across all four held-out partitions, higher predicted difficulty levels consistently corresponded to higher median observed non-anchor failure rates (Fig.~\ref{fig:rq1_boxplot}). In Verified, the median increased from 5.6\% at L1 through 14.1\%, 23.8\%, and 47.6\% at L2--L4 to 83.1\% at L5. The other three partitions showed the same monotonic increase across their populated levels.

The observed failure-rate distributions overlapped across levels.
Some partitions also had few or no tasks in the lower levels.
Nevertheless, the consistent increase in median failure rates across all four partitions indicates that the \aname{} difficulty model's predicted difficulty levels reflect meaningful differences in observed repair difficulty.

\begin{boxA}
\textbf{RQ1 Answer: }Across four held-out partitions, the \aname{} difficulty model consistently predicted relative task-difficulty rankings with moderate positive rank correlations and above-chance pairwise discrimination.
Moreover, median observed failure rates increased monotonically across the populated predicted difficulty levels (L1--L5), indicating that these levels reflect meaningful differences in observed repair difficulty.
\end{boxA}

%% file: sections/rq2.tex
\subsection{RQ2: How Do Feature-Group Contributions Vary with Predicted Task Difficulty?}
\label{sec:rq2}

RQ1 showed that the \aname{} difficulty model can predict repair-task difficulty on held-out benchmark partitions.
A difficulty score alone,
however, does not indicate what support a task needs. RQ2 therefore examines how the seven Feature Groups contribute to these predictions, testing whether salient groups differ across tasks, the premise of task-specific support (Section~\ref{sec:support}).  We ask (1) which groups contribute most across tasks and (2) whether the most salient group varies with predicted difficulty. We analyze 2,594 tasks from the four cross-partition holdouts in RQ1, using the feature-group contributions defined in Section~\ref{sec:dare} and the observed non-anchor difficulty defined in Section~\ref{sec:scale_alignment}.

\subsubsection{Global Feature-Group Importance}
\label{sec:rq2_global}

Four feature groups---Transformation ($T$), Validation ($V$), Specification--Solution Gap ($\mathit{SG}$), and Localization ($L$)---contributed most overall, but no single group dominated the difficulty model's predictions (Table~\ref{tab:rq2_importance}).
To quantify their relative importance, we average the absolute centered contribution of each group across tasks within each held-out partition. We then average these values equally across the four partitions and normalize the seven groups to sum to 100\%, yielding \emph{Feature-Group Importance}. Because this measure uses absolute contributions, it captures the magnitude of each group's contribution to the prediction score regardless of whether that contribution is positive or negative.

To examine the relationship with observed repair outcomes, we also compute Spearman's rank correlation ($\rho$) between each group's signed contribution and the observed failure rate among non-anchor agents. The correlations are likewise averaged equally across the four holdouts, and 95\% confidence intervals (CIs) are estimated using 2,000 task-level bootstrap repetitions (Table~\ref{tab:rq2_importance}).

\begin{table}[!t]
\centering
\caption{Feature-Group Importance and Spearman correlation with observed non-anchor difficulty.}
\label{tab:rq2_importance}
\footnotesize
\setlength{\tabcolsep}{2.5pt}
\begin{tabular}{@{}lcr@{}}
\toprule
Feature group & Importance (\%) & Spearman $\rho$ [95\% CI] \\
\midrule
Transformation ($T$) & 26.9 & .412 [.366, .455] \\
Validation ($V$) & 20.1 & .338 [.290, .380] \\
Specification--Solution Gap ($\mathit{SG}$) & 13.9 & .347 [.298, .393] \\
Localization ($L$) & 13.5 & .255 [.208, .301] \\
Context ($C$) & 11.0 & .030 [$-$.024, .082] \\
Specification ($S$) & 10.0 & .052 [.001, .105] \\
Interaction ($\mathit{INT}$) & 4.5 & .199 [.150, .247] \\
\bottomrule
\end{tabular}
\end{table}

$T$ had the highest importance (26.9\%) and the strongest correlation with observed difficulty ($\rho = .412$). $V$ (20.1\%), $\mathit{SG}$ (13.9\%), and $L$ (13.5\%) followed, giving these four groups a combined importance of 74.4\% (Table~\ref{tab:rq2_importance}). All four showed positive correlations with observed difficulty in every held-out partition (Supplementary Table S1 in~\cite{dare_replication}).

Even $T$, the most influential group, accounted for only 25.1--28.8\% of Feature-Group Importance within individual partitions (Supplementary Table S1 in~\cite{dare_replication}). This indicates that the difficulty model combines contributions from all seven feature groups to reflect different aspects of software repair (Section~\ref{sec:repair_task_metrics}; Table~\ref{tab:rq2_importance}). The predictive performance reported in RQ1 was also obtained using the difficulty model with all seven feature groups (Section~\ref{sec:model}; Section~\ref{sec:rq1}).

\subsubsection{Task-Level Salient Feature Groups}
\label{sec:rq2_salient}



\begin{figure}[!t]
\centering
\definecolor{rqheat0}{HTML}{F7FBFF}
\definecolor{rqheat1}{HTML}{F5FAFE}
\definecolor{rqheat2}{HTML}{F3F8FE}
\definecolor{rqheat3}{HTML}{F2F7FD}
\definecolor{rqheat4}{HTML}{EFF6FC}
\definecolor{rqheat5}{HTML}{EEF5FC}
\definecolor{rqheat6}{HTML}{EBF3FB}
\definecolor{rqheat7}{HTML}{EAF2FB}
\definecolor{rqheat8}{HTML}{E7F1FA}
\definecolor{rqheat9}{HTML}{E5EFF9}
\definecolor{rqheat10}{HTML}{E3EEF9}
\definecolor{rqheat11}{HTML}{E1EDF8}
\definecolor{rqheat12}{HTML}{DFECF7}
\definecolor{rqheat13}{HTML}{DDEAF7}
\definecolor{rqheat14}{HTML}{DCE9F6}
\definecolor{rqheat15}{HTML}{D9E8F5}
\definecolor{rqheat16}{HTML}{D8E7F5}
\definecolor{rqheat17}{HTML}{D6E5F4}
\definecolor{rqheat18}{HTML}{D3E4F3}
\definecolor{rqheat19}{HTML}{D2E3F3}
\definecolor{rqheat20}{HTML}{D0E1F2}
\definecolor{rqheat21}{HTML}{CEE0F2}
\definecolor{rqheat22}{HTML}{CCDFF1}
\definecolor{rqheat23}{HTML}{CADEF0}
\definecolor{rqheat24}{HTML}{C8DCF0}
\definecolor{rqheat25}{HTML}{C6DBEF}
\definecolor{rqheat26}{HTML}{C3DAEE}
\definecolor{rqheat27}{HTML}{BFD8ED}
\definecolor{rqheat29}{HTML}{B9D6EA}
\definecolor{rqheat30}{HTML}{B7D4EA}
\definecolor{rqheat31}{HTML}{B3D3E8}
\definecolor{rqheat33}{HTML}{ADD0E6}
\definecolor{rqheat37}{HTML}{A0CBE2}
\definecolor{rqheat38}{HTML}{9CC9E1}
\definecolor{rqheat40}{HTML}{94C4DF}
\definecolor{rqheat45}{HTML}{7FB9DA}
\definecolor{rqheat59}{HTML}{4D99CA}
\definecolor{rqheat61}{HTML}{4695C8}
\definecolor{rqheat62}{HTML}{4493C7}
\definecolor{rqheat100}{HTML}{08306B}
\definecolor{rqlegend0}{HTML}{F7FBFF}
\definecolor{rqlegend1}{HTML}{F5FAFE}
\definecolor{rqlegend2}{HTML}{F3F8FE}
\definecolor{rqlegend3}{HTML}{F2F7FD}
\definecolor{rqlegend4}{HTML}{EFF6FC}
\definecolor{rqlegend5}{HTML}{EEF5FC}
\definecolor{rqlegend6}{HTML}{EBF3FB}
\definecolor{rqlegend7}{HTML}{EAF2FB}
\definecolor{rqlegend8}{HTML}{E7F1FA}
\definecolor{rqlegend9}{HTML}{E5EFF9}
\definecolor{rqlegend10}{HTML}{E3EEF9}
\definecolor{rqlegend11}{HTML}{E1EDF8}
\definecolor{rqlegend12}{HTML}{DFECF7}
\definecolor{rqlegend13}{HTML}{DDEAF7}
\definecolor{rqlegend14}{HTML}{DCE9F6}
\definecolor{rqlegend15}{HTML}{D9E8F5}
\definecolor{rqlegend16}{HTML}{D8E7F5}
\definecolor{rqlegend17}{HTML}{D6E5F4}
\definecolor{rqlegend18}{HTML}{D3E4F3}
\definecolor{rqlegend19}{HTML}{D2E3F3}
\definecolor{rqlegend20}{HTML}{D0E1F2}
\definecolor{rqlegend21}{HTML}{CEE0F2}
\definecolor{rqlegend22}{HTML}{CCDFF1}
\definecolor{rqlegend23}{HTML}{CADEF0}
\definecolor{rqlegend24}{HTML}{C8DCF0}
\definecolor{rqlegend25}{HTML}{C6DBEF}
\definecolor{rqlegend26}{HTML}{C3DAEE}
\definecolor{rqlegend27}{HTML}{BFD8ED}
\definecolor{rqlegend28}{HTML}{BDD7EC}
\definecolor{rqlegend29}{HTML}{B9D6EA}
\definecolor{rqlegend30}{HTML}{B7D4EA}
\definecolor{rqlegend31}{HTML}{B3D3E8}
\definecolor{rqlegend32}{HTML}{B0D2E7}
\definecolor{rqlegend33}{HTML}{ADD0E6}
\definecolor{rqlegend34}{HTML}{A9CFE5}
\definecolor{rqlegend35}{HTML}{A6CEE4}
\definecolor{rqlegend36}{HTML}{A3CCE3}
\definecolor{rqlegend37}{HTML}{A0CBE2}
\definecolor{rqlegend38}{HTML}{9CC9E1}
\definecolor{rqlegend39}{HTML}{99C7E0}
\definecolor{rqlegend40}{HTML}{94C4DF}
\definecolor{rqlegend41}{HTML}{91C3DE}
\definecolor{rqlegend42}{HTML}{8CC0DD}
\definecolor{rqlegend43}{HTML}{87BDDC}
\definecolor{rqlegend44}{HTML}{84BCDB}
\definecolor{rqlegend45}{HTML}{7FB9DA}
\definecolor{rqlegend46}{HTML}{7CB7DA}
\definecolor{rqlegend47}{HTML}{77B5D9}
\definecolor{rqlegend48}{HTML}{74B3D8}
\definecolor{rqlegend49}{HTML}{6FB0D7}
\definecolor{rqlegend50}{HTML}{6AAED6}
\definecolor{rqlegend51}{HTML}{68ACD5}
\definecolor{rqlegend52}{HTML}{64A9D3}
\definecolor{rqlegend53}{HTML}{61A7D2}
\definecolor{rqlegend54}{HTML}{5DA5D1}
\definecolor{rqlegend55}{HTML}{5BA3D0}
\definecolor{rqlegend56}{HTML}{57A0CE}
\definecolor{rqlegend57}{HTML}{549FCD}
\definecolor{rqlegend58}{HTML}{519CCC}
\definecolor{rqlegend59}{HTML}{4D99CA}
\definecolor{rqlegend60}{HTML}{4A98C9}
\definecolor{rqlegend61}{HTML}{4695C8}
\definecolor{rqlegend62}{HTML}{4493C7}
\definecolor{rqlegend63}{HTML}{4090C5}
\definecolor{rqlegend64}{HTML}{3E8EC4}
\definecolor{rqlegend65}{HTML}{3B8BC2}
\definecolor{rqlegend66}{HTML}{3989C1}
\definecolor{rqlegend67}{HTML}{3686C0}
\definecolor{rqlegend68}{HTML}{3383BE}
\definecolor{rqlegend69}{HTML}{3181BD}
\definecolor{rqlegend70}{HTML}{2E7EBC}
\definecolor{rqlegend71}{HTML}{2C7CBA}
\definecolor{rqlegend72}{HTML}{2979B9}
\definecolor{rqlegend73}{HTML}{2676B8}
\definecolor{rqlegend74}{HTML}{2373B6}
\definecolor{rqlegend75}{HTML}{2070B4}
\definecolor{rqlegend76}{HTML}{1F6EB3}
\definecolor{rqlegend77}{HTML}{1C6BB0}
\definecolor{rqlegend78}{HTML}{1B69AF}
\definecolor{rqlegend79}{HTML}{1966AD}
\definecolor{rqlegend80}{HTML}{1764AB}
\definecolor{rqlegend81}{HTML}{1561A9}
\definecolor{rqlegend82}{HTML}{135FA7}
\definecolor{rqlegend83}{HTML}{115CA5}
\definecolor{rqlegend84}{HTML}{0E59A2}
\definecolor{rqlegend85}{HTML}{0D57A1}
\definecolor{rqlegend86}{HTML}{0A549E}
\definecolor{rqlegend87}{HTML}{09529D}
\definecolor{rqlegend88}{HTML}{084F99}
\definecolor{rqlegend89}{HTML}{084D96}
\definecolor{rqlegend90}{HTML}{084A91}
\definecolor{rqlegend91}{HTML}{08488E}
\definecolor{rqlegend92}{HTML}{08458A}
\definecolor{rqlegend93}{HTML}{084285}
\definecolor{rqlegend94}{HTML}{084082}
\definecolor{rqlegend95}{HTML}{083C7D}
\definecolor{rqlegend96}{HTML}{083A7A}
\definecolor{rqlegend97}{HTML}{083776}
\definecolor{rqlegend98}{HTML}{083573}
\definecolor{rqlegend99}{HTML}{08326E}
\definecolor{rqlegend100}{HTML}{08306B}
\resizebox{\columnwidth}{!}{%
\begin{tikzpicture}[x=1pt,y=-1pt,inner sep=0pt,outer sep=0pt]
  \path[use as bounding box] (0,0) rectangle (245,214);
  \node[font=\fontsize{7.2}{8.1}\selectfont\bfseries,anchor=center] at (70.000,9.000) {SWE-bench Verified};
  \node[font=\fontsize{6.0}{6.5}\selectfont,anchor=west] at (0.500,23.350) {L1};
  \node[font=\fontsize{4.2}{4.6}\selectfont,text=gray!75!black,anchor=east] at (24.000,23.350) {n=11};
  \filldraw[fill=rqheat9,draw=black!10,line width=.20pt] (25.400,17.000) rectangle (36.680,29.700);
  \node[font=\fontsize{6.0}{6.5}\selectfont,text=black!88,anchor=center] at (31.040,23.350) {9};
  \filldraw[fill=rqheat0,draw=black!10,line width=.20pt] (36.680,17.000) rectangle (47.960,29.700);
  \node[font=\fontsize{6.0}{6.5}\selectfont,text=black!88,anchor=center] at (42.320,23.350) {0};
  \filldraw[fill=rqheat0,draw=black!10,line width=.20pt] (47.960,17.000) rectangle (59.240,29.700);
  \node[font=\fontsize{6.0}{6.5}\selectfont,text=black!88,anchor=center] at (53.600,23.350) {0};
  \filldraw[fill=rqheat0,draw=black!10,line width=.20pt] (59.240,17.000) rectangle (70.520,29.700);
  \node[font=\fontsize{6.0}{6.5}\selectfont,text=black!88,anchor=center] at (64.880,23.350) {0};
  \filldraw[fill=rqheat0,draw=black!10,line width=.20pt] (70.520,17.000) rectangle (81.800,29.700);
  \node[font=\fontsize{6.0}{6.5}\selectfont,text=black!88,anchor=center] at (76.160,23.350) {0};
  \filldraw[fill=rqheat45,draw=black!10,line width=.20pt] (81.800,17.000) rectangle (93.080,29.700);
  \node[font=\fontsize{6.0}{6.5}\selectfont,text=black!88,anchor=center] at (87.440,23.350) {45};
  \filldraw[fill=rqheat18,draw=black!10,line width=.20pt] (93.080,17.000) rectangle (104.360,29.700);
  \node[font=\fontsize{6.0}{6.5}\selectfont,text=black!88,anchor=center] at (98.720,23.350) {18};
  \filldraw[fill=rqheat27,draw=black!10,line width=.20pt] (104.360,17.000) rectangle (115.640,29.700);
  \node[font=\fontsize{6.0}{6.5}\selectfont,text=black!88,anchor=center] at (110.000,23.350) {27};
  \node[font=\fontsize{6.0}{6.5}\selectfont,anchor=west] at (0.500,36.050) {L2};
  \node[font=\fontsize{4.2}{4.6}\selectfont,text=gray!75!black,anchor=east] at (24.000,36.050) {n=110};
  \filldraw[fill=rqheat27,draw=black!10,line width=.20pt] (25.400,29.700) rectangle (36.680,42.400);
  \node[font=\fontsize{6.0}{6.5}\selectfont,text=black!88,anchor=center] at (31.040,36.050) {27};
  \filldraw[fill=rqheat5,draw=black!10,line width=.20pt] (36.680,29.700) rectangle (47.960,42.400);
  \node[font=\fontsize{6.0}{6.5}\selectfont,text=black!88,anchor=center] at (42.320,36.050) {5};
  \filldraw[fill=rqheat8,draw=black!10,line width=.20pt] (47.960,29.700) rectangle (59.240,42.400);
  \node[font=\fontsize{6.0}{6.5}\selectfont,text=black!88,anchor=center] at (53.600,36.050) {8};
  \filldraw[fill=rqheat1,draw=black!10,line width=.20pt] (59.240,29.700) rectangle (70.520,42.400);
  \node[font=\fontsize{6.0}{6.5}\selectfont,text=black!88,anchor=center] at (64.880,36.050) {1};
  \filldraw[fill=rqheat5,draw=black!10,line width=.20pt] (70.520,29.700) rectangle (81.800,42.400);
  \node[font=\fontsize{6.0}{6.5}\selectfont,text=black!88,anchor=center] at (76.160,36.050) {5};
  \filldraw[fill=rqheat30,draw=black!10,line width=.20pt] (81.800,29.700) rectangle (93.080,42.400);
  \node[font=\fontsize{6.0}{6.5}\selectfont,text=black!88,anchor=center] at (87.440,36.050) {30};
  \filldraw[fill=rqheat14,draw=black!10,line width=.20pt] (93.080,29.700) rectangle (104.360,42.400);
  \node[font=\fontsize{6.0}{6.5}\selectfont,text=black!88,anchor=center] at (98.720,36.050) {14};
  \filldraw[fill=rqheat10,draw=black!10,line width=.20pt] (104.360,29.700) rectangle (115.640,42.400);
  \node[font=\fontsize{6.0}{6.5}\selectfont,text=black!88,anchor=center] at (110.000,36.050) {10};
  \node[font=\fontsize{6.0}{6.5}\selectfont,anchor=west] at (0.500,48.750) {L3};
  \node[font=\fontsize{4.2}{4.6}\selectfont,text=gray!75!black,anchor=east] at (24.000,48.750) {n=156};
  \filldraw[fill=rqheat24,draw=black!10,line width=.20pt] (25.400,42.400) rectangle (36.680,55.100);
  \node[font=\fontsize{6.0}{6.5}\selectfont,text=black!88,anchor=center] at (31.040,48.750) {24};
  \filldraw[fill=rqheat8,draw=black!10,line width=.20pt] (36.680,42.400) rectangle (47.960,55.100);
  \node[font=\fontsize{6.0}{6.5}\selectfont,text=black!88,anchor=center] at (42.320,48.750) {8};
  \filldraw[fill=rqheat23,draw=black!10,line width=.20pt] (47.960,42.400) rectangle (59.240,55.100);
  \node[font=\fontsize{6.0}{6.5}\selectfont,text=black!88,anchor=center] at (53.600,48.750) {23};
  \filldraw[fill=rqheat4,draw=black!10,line width=.20pt] (59.240,42.400) rectangle (70.520,55.100);
  \node[font=\fontsize{6.0}{6.5}\selectfont,text=black!88,anchor=center] at (64.880,48.750) {4};
  \filldraw[fill=rqheat13,draw=black!10,line width=.20pt] (70.520,42.400) rectangle (81.800,55.100);
  \node[font=\fontsize{6.0}{6.5}\selectfont,text=black!88,anchor=center] at (76.160,48.750) {13};
  \filldraw[fill=rqheat20,draw=black!10,line width=.20pt] (81.800,42.400) rectangle (93.080,55.100);
  \node[font=\fontsize{6.0}{6.5}\selectfont,text=black!88,anchor=center] at (87.440,48.750) {20};
  \filldraw[fill=rqheat6,draw=black!10,line width=.20pt] (93.080,42.400) rectangle (104.360,55.100);
  \node[font=\fontsize{6.0}{6.5}\selectfont,text=black!88,anchor=center] at (98.720,48.750) {6};
  \filldraw[fill=rqheat1,draw=black!10,line width=.20pt] (104.360,42.400) rectangle (115.640,55.100);
  \node[font=\fontsize{6.0}{6.5}\selectfont,text=black!88,anchor=center] at (110.000,48.750) {1};
  \node[font=\fontsize{6.0}{6.5}\selectfont,anchor=west] at (0.500,61.450) {L4};
  \node[font=\fontsize{4.2}{4.6}\selectfont,text=gray!75!black,anchor=east] at (24.000,61.450) {n=121};
  \filldraw[fill=rqheat13,draw=black!10,line width=.20pt] (25.400,55.100) rectangle (36.680,67.800);
  \node[font=\fontsize{6.0}{6.5}\selectfont,text=black!88,anchor=center] at (31.040,61.450) {13};
  \filldraw[fill=rqheat17,draw=black!10,line width=.20pt] (36.680,55.100) rectangle (47.960,67.800);
  \node[font=\fontsize{6.0}{6.5}\selectfont,text=black!88,anchor=center] at (42.320,61.450) {17};
  \filldraw[fill=rqheat33,draw=black!10,line width=.20pt] (47.960,55.100) rectangle (59.240,67.800);
  \node[font=\fontsize{6.0}{6.5}\selectfont,text=black!88,anchor=center] at (53.600,61.450) {33};
  \filldraw[fill=rqheat7,draw=black!10,line width=.20pt] (59.240,55.100) rectangle (70.520,67.800);
  \node[font=\fontsize{6.0}{6.5}\selectfont,text=black!88,anchor=center] at (64.880,61.450) {7};
  \filldraw[fill=rqheat17,draw=black!10,line width=.20pt] (70.520,55.100) rectangle (81.800,67.800);
  \node[font=\fontsize{6.0}{6.5}\selectfont,text=black!88,anchor=center] at (76.160,61.450) {17};
  \filldraw[fill=rqheat11,draw=black!10,line width=.20pt] (81.800,55.100) rectangle (93.080,67.800);
  \node[font=\fontsize{6.0}{6.5}\selectfont,text=black!88,anchor=center] at (87.440,61.450) {11};
  \filldraw[fill=rqheat2,draw=black!10,line width=.20pt] (93.080,55.100) rectangle (104.360,67.800);
  \node[font=\fontsize{6.0}{6.5}\selectfont,text=black!88,anchor=center] at (98.720,61.450) {2};
  \filldraw[fill=rqheat0,draw=black!10,line width=.20pt] (104.360,55.100) rectangle (115.640,67.800);
  \node[font=\fontsize{6.0}{6.5}\selectfont,text=black!88,anchor=center] at (110.000,61.450) {0};
  \node[font=\fontsize{6.0}{6.5}\selectfont,anchor=west] at (0.500,74.150) {L5};
  \node[font=\fontsize{4.2}{4.6}\selectfont,text=gray!75!black,anchor=east] at (24.000,74.150) {n=102};
  \filldraw[fill=rqheat8,draw=black!10,line width=.20pt] (25.400,67.800) rectangle (36.680,80.500);
  \node[font=\fontsize{6.0}{6.5}\selectfont,text=black!88,anchor=center] at (31.040,74.150) {8};
  \filldraw[fill=rqheat25,draw=black!10,line width=.20pt] (36.680,67.800) rectangle (47.960,80.500);
  \node[font=\fontsize{6.0}{6.5}\selectfont,text=black!88,anchor=center] at (42.320,74.150) {25};
  \filldraw[fill=rqheat17,draw=black!10,line width=.20pt] (47.960,67.800) rectangle (59.240,80.500);
  \node[font=\fontsize{6.0}{6.5}\selectfont,text=black!88,anchor=center] at (53.600,74.150) {17};
  \filldraw[fill=rqheat37,draw=black!10,line width=.20pt] (59.240,67.800) rectangle (70.520,80.500);
  \node[font=\fontsize{6.0}{6.5}\selectfont,text=black!88,anchor=center] at (64.880,74.150) {37};
  \filldraw[fill=rqheat13,draw=black!10,line width=.20pt] (70.520,67.800) rectangle (81.800,80.500);
  \node[font=\fontsize{6.0}{6.5}\selectfont,text=black!88,anchor=center] at (76.160,74.150) {13};
  \filldraw[fill=rqheat0,draw=black!10,line width=.20pt] (81.800,67.800) rectangle (93.080,80.500);
  \node[font=\fontsize{6.0}{6.5}\selectfont,text=black!88,anchor=center] at (87.440,74.150) {0};
  \filldraw[fill=rqheat0,draw=black!10,line width=.20pt] (93.080,67.800) rectangle (104.360,80.500);
  \node[font=\fontsize{6.0}{6.5}\selectfont,text=black!88,anchor=center] at (98.720,74.150) {0};
  \filldraw[fill=rqheat0,draw=black!10,line width=.20pt] (104.360,67.800) rectangle (115.640,80.500);
  \node[font=\fontsize{6.0}{6.5}\selectfont,text=black!88,anchor=center] at (110.000,74.150) {0};
  \node[font=\fontsize{5.7}{6.2}\selectfont,anchor=north] at (31.040,82.500) {S};
  \node[font=\fontsize{5.7}{6.2}\selectfont,anchor=north] at (42.320,82.500) {SG};
  \node[font=\fontsize{5.7}{6.2}\selectfont,anchor=north] at (53.600,82.500) {L};
  \node[font=\fontsize{5.7}{6.2}\selectfont,anchor=north] at (64.880,82.500) {T};
  \node[font=\fontsize{5.7}{6.2}\selectfont,anchor=north] at (76.160,82.500) {V};
  \node[font=\fontsize{5.7}{6.2}\selectfont,anchor=north] at (87.440,82.500) {C};
  \node[font=\fontsize{5.7}{6.2}\selectfont,anchor=north] at (97.720,82.500) {INT};
  \node[font=\fontsize{5.7}{6.2}\selectfont,anchor=north] at (111.000,82.500) {None};
  \foreach \y in {29.700,42.400,55.100,67.800}
    \draw[draw=black,line width=.30pt] (25.400,\y) -- (115.640,\y);
  \draw[draw=black!55,line width=.38pt] (25.400,17.000) rectangle (115.640,80.500);
  \node[font=\fontsize{7.2}{8.1}\selectfont\bfseries,anchor=center] at (193.000,9.000) {Lite $\setminus$ Verified};
  \node[font=\fontsize{6.0}{6.5}\selectfont,anchor=west] at (123.500,23.350) {L1};
  \node[font=\fontsize{4.2}{4.6}\selectfont,text=gray!75!black,anchor=east] at (147.000,23.350) {n=0};
  \filldraw[fill=black!5,draw=black!10,line width=.20pt] 
  (148.400,17.000) rectangle (159.680,29.700);
  \filldraw[fill=black!5,draw=black!10,line width=.20pt] (159.680,17.000) rectangle (170.960,29.700);
  \filldraw[fill=black!5,draw=black!10,line width=.20pt] (170.960,17.000) rectangle (182.240,29.700);
  \filldraw[fill=black!5,draw=black!10,line width=.20pt] (182.240,17.000) rectangle (193.520,29.700);
  \filldraw[fill=black!5,draw=black!10,line width=.20pt] (193.520,17.000) rectangle (204.800,29.700);
  \filldraw[fill=black!5,draw=black!10,line width=.20pt] (204.800,17.000) rectangle (216.080,29.700);
  \filldraw[fill=black!5,draw=black!10,line width=.20pt] (216.080,17.000) rectangle (227.360,29.700);
  \filldraw[fill=black!5,draw=black!10,line width=.20pt] (227.360,17.000) rectangle (238.640,29.700);
  \node[font=\fontsize{6.0}{6.5}\selectfont,anchor=west] at (123.500,36.050) {L2};
  \node[font=\fontsize{4.2}{4.6}\selectfont,text=gray!75!black,anchor=east] at (147.000,36.050) {n=8};
  \filldraw[fill=rqheat25,draw=black!10,line width=.20pt] (148.400,29.700) rectangle (159.680,42.400);
  \node[font=\fontsize{6.0}{6.5}\selectfont,text=black!88,anchor=center] at (154.040,36.050) {25};
  \filldraw[fill=rqheat13,draw=black!10,line width=.20pt] (159.680,29.700) rectangle (170.960,42.400);
  \node[font=\fontsize{6.0}{6.5}\selectfont,text=black!88,anchor=center] at (165.320,36.050) {13};
  \filldraw[fill=rqheat0,draw=black!10,line width=.20pt] (170.960,29.700) rectangle (182.240,42.400);
  \node[font=\fontsize{6.0}{6.5}\selectfont,text=black!88,anchor=center] at (176.600,36.050) {0};
  \filldraw[fill=rqheat0,draw=black!10,line width=.20pt] (182.240,29.700) rectangle (193.520,42.400);
  \node[font=\fontsize{6.0}{6.5}\selectfont,text=black!88,anchor=center] at (187.880,36.050) {0};
  \filldraw[fill=rqheat13,draw=black!10,line width=.20pt] (193.520,29.700) rectangle (204.800,42.400);
  \node[font=\fontsize{6.0}{6.5}\selectfont,text=black!88,anchor=center] at (199.160,36.050) {13};
  \filldraw[fill=rqheat25,draw=black!10,line width=.20pt] (204.800,29.700) rectangle (216.080,42.400);
  \node[font=\fontsize{6.0}{6.5}\selectfont,text=black!88,anchor=center] at (210.440,36.050) {25};
  \filldraw[fill=rqheat0,draw=black!10,line width=.20pt] (216.080,29.700) rectangle (227.360,42.400);
  \node[font=\fontsize{6.0}{6.5}\selectfont,text=black!88,anchor=center] at (221.720,36.050) {0};
  \filldraw[fill=rqheat25,draw=black!10,line width=.20pt] (227.360,29.700) rectangle (238.640,42.400);
  \node[font=\fontsize{6.0}{6.5}\selectfont,text=black!88,anchor=center] at (233.000,36.050) {25};
  \node[font=\fontsize{6.0}{6.5}\selectfont,anchor=west] at (123.500,48.750) {L3};
  \node[font=\fontsize{4.2}{4.6}\selectfont,text=gray!75!black,anchor=east] at (147.000,48.750) {n=28};
  \filldraw[fill=rqheat11,draw=black!10,line width=.20pt] (148.400,42.400) rectangle (159.680,55.100);
  \node[font=\fontsize{6.0}{6.5}\selectfont,text=black!88,anchor=center] at (154.040,48.750) {11};
  \filldraw[fill=rqheat7,draw=black!10,line width=.20pt] (159.680,42.400) rectangle (170.960,55.100);
  \node[font=\fontsize{6.0}{6.5}\selectfont,text=black!88,anchor=center] at (165.320,48.750) {7};
  \filldraw[fill=rqheat25,draw=black!10,line width=.20pt] (170.960,42.400) rectangle (182.240,55.100);
  \node[font=\fontsize{6.0}{6.5}\selectfont,text=black!88,anchor=center] at (176.600,48.750) {25};
  \filldraw[fill=rqheat0,draw=black!10,line width=.20pt] (182.240,42.400) rectangle (193.520,55.100);
  \node[font=\fontsize{6.0}{6.5}\selectfont,text=black!88,anchor=center] at (187.880,48.750) {0};
  \filldraw[fill=rqheat14,draw=black!10,line width=.20pt] (193.520,42.400) rectangle (204.800,55.100);
  \node[font=\fontsize{6.0}{6.5}\selectfont,text=black!88,anchor=center] at (199.160,48.750) {14};
  \filldraw[fill=rqheat21,draw=black!10,line width=.20pt] (204.800,42.400) rectangle (216.080,55.100);
  \node[font=\fontsize{6.0}{6.5}\selectfont,text=black!88,anchor=center] at (210.440,48.750) {21};
  \filldraw[fill=rqheat11,draw=black!10,line width=.20pt] (216.080,42.400) rectangle (227.360,55.100);
  \node[font=\fontsize{6.0}{6.5}\selectfont,text=black!88,anchor=center] at (221.720,48.750) {11};
  \filldraw[fill=rqheat11,draw=black!10,line width=.20pt] (227.360,42.400) rectangle (238.640,55.100);
  \node[font=\fontsize{6.0}{6.5}\selectfont,text=black!88,anchor=center] at (233.000,48.750) {11};
  \node[font=\fontsize{6.0}{6.5}\selectfont,anchor=west] at (123.500,61.450) {L4};
  \node[font=\fontsize{4.2}{4.6}\selectfont,text=gray!75!black,anchor=east] at (147.000,61.450) {n=70};
  \filldraw[fill=rqheat19,draw=black!10,line width=.20pt] (148.400,55.100) rectangle (159.680,67.800);
  \node[font=\fontsize{6.0}{6.5}\selectfont,text=black!88,anchor=center] at (154.040,61.450) {19};
  \filldraw[fill=rqheat19,draw=black!10,line width=.20pt] (159.680,55.100) rectangle (170.960,67.800);
  \node[font=\fontsize{6.0}{6.5}\selectfont,text=black!88,anchor=center] at (165.320,61.450) {19};
  \filldraw[fill=rqheat21,draw=black!10,line width=.20pt] (170.960,55.100) rectangle (182.240,67.800);
  \node[font=\fontsize{6.0}{6.5}\selectfont,text=black!88,anchor=center] at (176.600,61.450) {21};
  \filldraw[fill=rqheat10,draw=black!10,line width=.20pt] (182.240,55.100) rectangle (193.520,67.800);
  \node[font=\fontsize{6.0}{6.5}\selectfont,text=black!88,anchor=center] at (187.880,61.450) {10};
  \filldraw[fill=rqheat10,draw=black!10,line width=.20pt] (193.520,55.100) rectangle (204.800,67.800);
  \node[font=\fontsize{6.0}{6.5}\selectfont,text=black!88,anchor=center] at (199.160,61.450) {10};
  \filldraw[fill=rqheat9,draw=black!10,line width=.20pt] (204.800,55.100) rectangle (216.080,67.800);
  \node[font=\fontsize{6.0}{6.5}\selectfont,text=black!88,anchor=center] at (210.440,61.450) {9};
  \filldraw[fill=rqheat9,draw=black!10,line width=.20pt] (216.080,55.100) rectangle (227.360,67.800);
  \node[font=\fontsize{6.0}{6.5}\selectfont,text=black!88,anchor=center] at (221.720,61.450) {9};
  \filldraw[fill=rqheat4,draw=black!10,line width=.20pt] (227.360,55.100) rectangle (238.640,67.800);
  \node[font=\fontsize{6.0}{6.5}\selectfont,text=black!88,anchor=center] at (233.000,61.450) {4};
  \node[font=\fontsize{6.0}{6.5}\selectfont,anchor=west] at (123.500,74.150) {L5};
  \node[font=\fontsize{4.2}{4.6}\selectfont,text=gray!75!black,anchor=east] at (147.000,74.150) {n=101};
  \filldraw[fill=rqheat8,draw=black!10,line width=.20pt] (148.400,67.800) rectangle (159.680,80.500);
  \node[font=\fontsize{6.0}{6.5}\selectfont,text=black!88,anchor=center] at (154.040,74.150) {8};
  \filldraw[fill=rqheat29,draw=black!10,line width=.20pt] (159.680,67.800) rectangle (170.960,80.500);
  \node[font=\fontsize{6.0}{6.5}\selectfont,text=black!88,anchor=center] at (165.320,74.150) {29};
  \filldraw[fill=rqheat21,draw=black!10,line width=.20pt] (170.960,67.800) rectangle (182.240,80.500);
  \node[font=\fontsize{6.0}{6.5}\selectfont,text=black!88,anchor=center] at (176.600,74.150) {21};
  \filldraw[fill=rqheat19,draw=black!10,line width=.20pt] (182.240,67.800) rectangle (193.520,80.500);
  \node[font=\fontsize{6.0}{6.5}\selectfont,text=black!88,anchor=center] at (187.880,74.150) {19};
  \filldraw[fill=rqheat22,draw=black!10,line width=.20pt] (193.520,67.800) rectangle (204.800,80.500);
  \node[font=\fontsize{6.0}{6.5}\selectfont,text=black!88,anchor=center] at (199.160,74.150) {22};
  \filldraw[fill=rqheat2,draw=black!10,line width=.20pt] (204.800,67.800) rectangle (216.080,80.500);
  \node[font=\fontsize{6.0}{6.5}\selectfont,text=black!88,anchor=center] at (210.440,74.150) {2};
  \filldraw[fill=rqheat0,draw=black!10,line width=.20pt] (216.080,67.800) rectangle (227.360,80.500);
  \node[font=\fontsize{6.0}{6.5}\selectfont,text=black!88,anchor=center] at (221.720,74.150) {0};
  \filldraw[fill=rqheat0,draw=black!10,line width=.20pt] (227.360,67.800) rectangle (238.640,80.500);
  \node[font=\fontsize{6.0}{6.5}\selectfont,text=black!88,anchor=center] at (233.000,74.150) {0};
  \node[font=\fontsize{5.7}{6.2}\selectfont,anchor=north] at (154.040,82.500) {S};
  \node[font=\fontsize{5.7}{6.2}\selectfont,anchor=north] at (165.320,82.500) {SG};
  \node[font=\fontsize{5.7}{6.2}\selectfont,anchor=north] at (176.600,82.500) {L};
  \node[font=\fontsize{5.7}{6.2}\selectfont,anchor=north] at (187.880,82.500) {T};
  \node[font=\fontsize{5.7}{6.2}\selectfont,anchor=north] at (199.160,82.500) {V};
  \node[font=\fontsize{5.7}{6.2}\selectfont,anchor=north] at (210.440,82.500) {C};
  \node[font=\fontsize{5.7}{6.2}\selectfont,anchor=north] at (220.720,82.500) {INT};
  \node[font=\fontsize{5.7}{6.2}\selectfont,anchor=north] at (234.000,82.500) {None};
   \foreach \y in {29.700,42.400,55.100,67.800}
    \draw[draw=black,line width=.30pt] (148.400,\y) -- (238.640,\y);
  \draw[draw=black!55,line width=.38pt] (148.400,17.000) rectangle (238.640,80.500);
  \node[font=\fontsize{7.2}{8.1}\selectfont\bfseries,anchor=center] at (70.000,102.000) {Full $\setminus$ (Verified $\cup$ Lite)};
  \node[font=\fontsize{6.0}{6.5}\selectfont,anchor=west] at (0.500,116.350) {L1};
  \node[font=\fontsize{4.2}{4.6}\selectfont,text=gray!75!black,anchor=east] at (24.000,116.350) {n=0};
  \filldraw[fill=black!5,draw=black!10,line width=.20pt] (25.400,110.000) rectangle (36.680,122.700);
  \filldraw[fill=black!5,draw=black!10,line width=.20pt] (36.680,110.000) rectangle (47.960,122.700);
  \filldraw[fill=black!5,draw=black!10,line width=.20pt] (47.960,110.000) rectangle (59.240,122.700);
  \filldraw[fill=black!5,draw=black!10,line width=.20pt] (59.240,110.000) rectangle (70.520,122.700);
  \filldraw[fill=black!5,draw=black!10,line width=.20pt] (70.520,110.000) rectangle (81.800,122.700);
  \filldraw[fill=black!5,draw=black!10,line width=.20pt] (81.800,110.000) rectangle (93.080,122.700);
  \filldraw[fill=black!5,draw=black!10,line width=.20pt] (93.080,110.000) rectangle (104.360,122.700);
  \filldraw[fill=black!5,draw=black!10,line width=.20pt] (104.360,110.000) rectangle (115.640,122.700);
  \node[font=\fontsize{6.0}{6.5}\selectfont,anchor=west] at (0.500,129.050) {L2};
  \node[font=\fontsize{4.2}{4.6}\selectfont,text=gray!75!black,anchor=east] at (24.000,129.050) {n=17};
  \filldraw[fill=rqheat24,draw=black!10,line width=.20pt] (25.400,122.700) rectangle (36.680,135.400);
  \node[font=\fontsize{6.0}{6.5}\selectfont,text=black!88,anchor=center] at (31.040,129.050) {24};
  \filldraw[fill=rqheat0,draw=black!10,line width=.20pt] (36.680,122.700) rectangle (47.960,135.400);
  \node[font=\fontsize{6.0}{6.5}\selectfont,text=black!88,anchor=center] at (42.320,129.050) {0};
  \filldraw[fill=rqheat0,draw=black!10,line width=.20pt] (47.960,122.700) rectangle (59.240,135.400);
  \node[font=\fontsize{6.0}{6.5}\selectfont,text=black!88,anchor=center] at (53.600,129.050) {0};
  \filldraw[fill=rqheat0,draw=black!10,line width=.20pt] (59.240,122.700) rectangle (70.520,135.400);
  \node[font=\fontsize{6.0}{6.5}\selectfont,text=black!88,anchor=center] at (64.880,129.050) {0};
  \filldraw[fill=rqheat0,draw=black!10,line width=.20pt] (70.520,122.700) rectangle (81.800,135.400);
  \node[font=\fontsize{6.0}{6.5}\selectfont,text=black!88,anchor=center] at (76.160,129.050) {0};
  \filldraw[fill=rqheat59,draw=black!10,line width=.20pt] (81.800,122.700) rectangle (93.080,135.400);
  \node[font=\fontsize{6.0}{6.5}\selectfont,text=white,anchor=center] at (87.440,129.050) {59};
  \filldraw[fill=rqheat0,draw=black!10,line width=.20pt] (93.080,122.700) rectangle (104.360,135.400);
  \node[font=\fontsize{6.0}{6.5}\selectfont,text=black!88,anchor=center] at (98.720,129.050) {0};
  \filldraw[fill=rqheat18,draw=black!10,line width=.20pt] (104.360,122.700) rectangle (115.640,135.400);
  \node[font=\fontsize{6.0}{6.5}\selectfont,text=black!88,anchor=center] at (110.000,129.050) {18};
  \node[font=\fontsize{6.0}{6.5}\selectfont,anchor=west] at (0.500,141.750) {L3};
  \node[font=\fontsize{4.2}{4.6}\selectfont,text=gray!75!black,anchor=east] at (24.000,141.750) {n=103};
  \filldraw[fill=rqheat19,draw=black!10,line width=.20pt] (25.400,135.400) rectangle (36.680,148.100);
  \node[font=\fontsize{6.0}{6.5}\selectfont,text=black!88,anchor=center] at (31.040,141.750) {19};
  \filldraw[fill=rqheat2,draw=black!10,line width=.20pt] (36.680,135.400) rectangle (47.960,148.100);
  \node[font=\fontsize{6.0}{6.5}\selectfont,text=black!88,anchor=center] at (42.320,141.750) {2};
  \filldraw[fill=rqheat3,draw=black!10,line width=.20pt] (47.960,135.400) rectangle (59.240,148.100);
  \node[font=\fontsize{6.0}{6.5}\selectfont,text=black!88,anchor=center] at (53.600,141.750) {3};
  \filldraw[fill=rqheat1,draw=black!10,line width=.20pt] (59.240,135.400) rectangle (70.520,148.100);
  \node[font=\fontsize{6.0}{6.5}\selectfont,text=black!88,anchor=center] at (64.880,141.750) {1};
  \filldraw[fill=rqheat12,draw=black!10,line width=.20pt] (70.520,135.400) rectangle (81.800,148.100);
  \node[font=\fontsize{6.0}{6.5}\selectfont,text=black!88,anchor=center] at (76.160,141.750) {12};
  \filldraw[fill=rqheat61,draw=black!10,line width=.20pt] (81.800,135.400) rectangle (93.080,148.100);
  \node[font=\fontsize{6.0}{6.5}\selectfont,text=white,anchor=center] at (87.440,141.750) {61};
  \filldraw[fill=rqheat1,draw=black!10,line width=.20pt] (93.080,135.400) rectangle (104.360,148.100);
  \node[font=\fontsize{6.0}{6.5}\selectfont,text=black!88,anchor=center] at (98.720,141.750) {1};
  \filldraw[fill=rqheat1,draw=black!10,line width=.20pt] (104.360,135.400) rectangle (115.640,148.100);
  \node[font=\fontsize{6.0}{6.5}\selectfont,text=black!88,anchor=center] at (110.000,141.750) {1};
  \node[font=\fontsize{6.0}{6.5}\selectfont,anchor=west] at (0.500,154.450) {L4};
  \node[font=\fontsize{4.2}{4.6}\selectfont,text=gray!75!black,anchor=east] at (24.000,154.450) {n=287};
  \filldraw[fill=rqheat16,draw=black!10,line width=.20pt] (25.400,148.100) rectangle (36.680,160.800);
  \node[font=\fontsize{6.0}{6.5}\selectfont,text=black!88,anchor=center] at (31.040,154.450) {16};
  \filldraw[fill=rqheat8,draw=black!10,line width=.20pt] (36.680,148.100) rectangle (47.960,160.800);
  \node[font=\fontsize{6.0}{6.5}\selectfont,text=black!88,anchor=center] at (42.320,154.450) {8};
  \filldraw[fill=rqheat11,draw=black!10,line width=.20pt] (47.960,148.100) rectangle (59.240,160.800);
  \node[font=\fontsize{6.0}{6.5}\selectfont,text=black!88,anchor=center] at (53.600,154.450) {11};
  \filldraw[fill=rqheat7,draw=black!10,line width=.20pt] (59.240,148.100) rectangle (70.520,160.800);
  \node[font=\fontsize{6.0}{6.5}\selectfont,text=black!88,anchor=center] at (64.880,154.450) {7};
  \filldraw[fill=rqheat18,draw=black!10,line width=.20pt] (70.520,148.100) rectangle (81.800,160.800);
  \node[font=\fontsize{6.0}{6.5}\selectfont,text=black!88,anchor=center] at (76.160,154.450) {18};
  \filldraw[fill=rqheat38,draw=black!10,line width=.20pt] (81.800,148.100) rectangle (93.080,160.800);
  \node[font=\fontsize{6.0}{6.5}\selectfont,text=black!88,anchor=center] at (87.440,154.450) {38};
  \filldraw[fill=rqheat1,draw=black!10,line width=.20pt] (93.080,148.100) rectangle (104.360,160.800);
  \node[font=\fontsize{6.0}{6.5}\selectfont,text=black!88,anchor=center] at (98.720,154.450) {1};
  \filldraw[fill=rqheat0,draw=black!10,line width=.20pt] (104.360,148.100) rectangle (115.640,160.800);
  \node[font=\fontsize{6.0}{6.5}\selectfont,text=black!88,anchor=center] at (110.000,154.450) {0};
  \node[font=\fontsize{6.0}{6.5}\selectfont,anchor=west] at (0.500,167.150) {L5};
  \node[font=\fontsize{4.2}{4.6}\selectfont,text=gray!75!black,anchor=east] at (24.000,167.150) {n=1,180};
  \filldraw[fill=rqheat8,draw=black!10,line width=.20pt] (25.400,160.800) rectangle (36.680,173.500);
  \node[font=\fontsize{6.0}{6.5}\selectfont,text=black!88,anchor=center] at (31.040,167.150) {8};
  \filldraw[fill=rqheat4,draw=black!10,line width=.20pt] (36.680,160.800) rectangle (47.960,173.500);
  \node[font=\fontsize{6.0}{6.5}\selectfont,text=black!88,anchor=center] at (42.320,167.150) {4};
  \filldraw[fill=rqheat12,draw=black!10,line width=.20pt] (47.960,160.800) rectangle (59.240,173.500);
  \node[font=\fontsize{6.0}{6.5}\selectfont,text=black!88,anchor=center] at (53.600,167.150) {12};
  \filldraw[fill=rqheat40,draw=black!10,line width=.20pt] (59.240,160.800) rectangle (70.520,173.500);
  \node[font=\fontsize{6.0}{6.5}\selectfont,text=black!88,anchor=center] at (64.880,167.150) {40};
  \filldraw[fill=rqheat29,draw=black!10,line width=.20pt] (70.520,160.800) rectangle (81.800,173.500);
  \node[font=\fontsize{6.0}{6.5}\selectfont,text=black!88,anchor=center] at (76.160,167.150) {29};
  \filldraw[fill=rqheat6,draw=black!10,line width=.20pt] (81.800,160.800) rectangle (93.080,173.500);
  \node[font=\fontsize{6.0}{6.5}\selectfont,text=black!88,anchor=center] at (87.440,167.150) {6};
  \filldraw[fill=rqheat0,draw=black!10,line width=.20pt] (93.080,160.800) rectangle (104.360,173.500);
  \node[font=\fontsize{6.0}{6.5}\selectfont,text=black!88,anchor=center] at (98.720,167.150) {0};
  \filldraw[fill=rqheat0,draw=black!10,line width=.20pt] (104.360,160.800) rectangle (115.640,173.500);
  \node[font=\fontsize{6.0}{6.5}\selectfont,text=black!88,anchor=center] at (110.000,167.150) {0};
  \node[font=\fontsize{5.7}{6.2}\selectfont,anchor=north] at (31.040,175.500) {S};
  \node[font=\fontsize{5.7}{6.2}\selectfont,anchor=north] at (42.320,175.500) {SG};
  \node[font=\fontsize{5.7}{6.2}\selectfont,anchor=north] at (53.600,175.500) {L};
  \node[font=\fontsize{5.7}{6.2}\selectfont,anchor=north] at (64.880,175.500) {T};
  \node[font=\fontsize{5.7}{6.2}\selectfont,anchor=north] at (76.160,175.500) {V};
  \node[font=\fontsize{5.7}{6.2}\selectfont,anchor=north] at (87.440,175.500) {C};
  \node[font=\fontsize{5.7}{6.2}\selectfont,anchor=north] at (97.720,175.500) {INT};
  \node[font=\fontsize{5.7}{6.2}\selectfont,anchor=north] at (111.000,175.500) {None};
  \foreach \y in {122.700,135.400,148.100,160.800}
    \draw[draw=black,line width=.30pt] (25.400,\y) -- (115.640,\y);
  \draw[draw=black!55,line width=.38pt] (25.400,110.000) rectangle (115.640,173.500);
  \node[font=\fontsize{7.2}{8.1}\selectfont\bfseries,anchor=center] at (193.000,102.000) {SWE-bench-Live Lite};
  \node[font=\fontsize{6.0}{6.5}\selectfont,anchor=west] at (123.500,116.350) {L1};
  \node[font=\fontsize{4.2}{4.6}\selectfont,text=gray!75!black,anchor=east] at (147.000,116.350) {n=0};
  \filldraw[fill=black!5,draw=black!10,line width=.20pt] (148.400,110.000) rectangle (159.680,122.700);
  \filldraw[fill=black!5,draw=black!10,line width=.20pt] (159.680,110.000) rectangle (170.960,122.700);
  \filldraw[fill=black!5,draw=black!10,line width=.20pt] (170.960,110.000) rectangle (182.240,122.700);
  \filldraw[fill=black!5,draw=black!10,line width=.20pt] (182.240,110.000) rectangle (193.520,122.700);
  \filldraw[fill=black!5,draw=black!10,line width=.20pt] (193.520,110.000) rectangle (204.800,122.700);
  \filldraw[fill=black!5,draw=black!10,line width=.20pt] (204.800,110.000) rectangle (216.080,122.700);
  \filldraw[fill=black!5,draw=black!10,line width=.20pt] (216.080,110.000) rectangle (227.360,122.700);
  \filldraw[fill=black!5,draw=black!10,line width=.20pt] (227.360,110.000) rectangle (238.640,122.700);
  \node[font=\fontsize{6.0}{6.5}\selectfont,anchor=west] at (123.500,129.050) {L2};
  \node[font=\fontsize{4.2}{4.6}\selectfont,text=gray!75!black,anchor=east] at (147.000,129.050) {n=1};
  \filldraw[fill=rqheat0,draw=black!10,line width=.20pt] (148.400,122.700) rectangle (159.680,135.400);
  \node[font=\fontsize{6.0}{6.5}\selectfont,text=black!88,anchor=center] at (154.040,129.050) {0};
  \filldraw[fill=rqheat0,draw=black!10,line width=.20pt] (159.680,122.700) rectangle (170.960,135.400);
  \node[font=\fontsize{6.0}{6.5}\selectfont,text=black!88,anchor=center] at (165.320,129.050) {0};
  \filldraw[fill=rqheat0,draw=black!10,line width=.20pt] (170.960,122.700) rectangle (182.240,135.400);
  \node[font=\fontsize{6.0}{6.5}\selectfont,text=black!88,anchor=center] at (176.600,129.050) {0};
  \filldraw[fill=rqheat0,draw=black!10,line width=.20pt] (182.240,122.700) rectangle (193.520,135.400);
  \node[font=\fontsize{6.0}{6.5}\selectfont,text=black!88,anchor=center] at (187.880,129.050) {0};
  \filldraw[fill=rqheat0,draw=black!10,line width=.20pt] (193.520,122.700) rectangle (204.800,135.400);
  \node[font=\fontsize{6.0}{6.5}\selectfont,text=black!88,anchor=center] at (199.160,129.050) {0};
  \filldraw[fill=rqheat0,draw=black!10,line width=.20pt] (204.800,122.700) rectangle (216.080,135.400);
  \node[font=\fontsize{6.0}{6.5}\selectfont,text=black!88,anchor=center] at (210.440,129.050) {0};
  \filldraw[fill=rqheat0,draw=black!10,line width=.20pt] (216.080,122.700) rectangle (227.360,135.400);
  \node[font=\fontsize{6.0}{6.5}\selectfont,text=black!88,anchor=center] at (221.720,129.050) {0};
  \filldraw[fill=rqheat100,draw=black!10,line width=.20pt] (227.360,122.700) rectangle (238.640,135.400);
  \node[font=\fontsize{6.0}{6.5}\selectfont,text=white,anchor=center] at (233.000,129.050) {100};
  \node[font=\fontsize{6.0}{6.5}\selectfont,anchor=west] at (123.500,141.750) {L3};
  \node[font=\fontsize{4.2}{4.6}\selectfont,text=gray!75!black,anchor=east] at (147.000,141.750) {n=13};
  \filldraw[fill=rqheat38,draw=black!10,line width=.20pt] (148.400,135.400) rectangle (159.680,148.100);
  \node[font=\fontsize{6.0}{6.5}\selectfont,text=black!88,anchor=center] at (154.040,141.750) {38};
  \filldraw[fill=rqheat0,draw=black!10,line width=.20pt] (159.680,135.400) rectangle (170.960,148.100);
  \node[font=\fontsize{6.0}{6.5}\selectfont,text=black!88,anchor=center] at (165.320,141.750) {0};
  \filldraw[fill=rqheat23,draw=black!10,line width=.20pt] (170.960,135.400) rectangle (182.240,148.100);
  \node[font=\fontsize{6.0}{6.5}\selectfont,text=black!88,anchor=center] at (176.600,141.750) {23};
  \filldraw[fill=rqheat8,draw=black!10,line width=.20pt] (182.240,135.400) rectangle (193.520,148.100);
  \node[font=\fontsize{6.0}{6.5}\selectfont,text=black!88,anchor=center] at (187.880,141.750) {8};
  \filldraw[fill=rqheat0,draw=black!10,line width=.20pt] (193.520,135.400) rectangle (204.800,148.100);
  \node[font=\fontsize{6.0}{6.5}\selectfont,text=black!88,anchor=center] at (199.160,141.750) {0};
  \filldraw[fill=rqheat0,draw=black!10,line width=.20pt] (204.800,135.400) rectangle (216.080,148.100);
  \node[font=\fontsize{6.0}{6.5}\selectfont,text=black!88,anchor=center] at (210.440,141.750) {0};
  \filldraw[fill=rqheat0,draw=black!10,line width=.20pt] (216.080,135.400) rectangle (227.360,148.100);
  \node[font=\fontsize{6.0}{6.5}\selectfont,text=black!88,anchor=center] at (221.720,141.750) {0};
  \filldraw[fill=rqheat31,draw=black!10,line width=.20pt] (227.360,135.400) rectangle (238.640,148.100);
  \node[font=\fontsize{6.0}{6.5}\selectfont,text=black!88,anchor=center] at (233.000,141.750) {31};
  \node[font=\fontsize{6.0}{6.5}\selectfont,anchor=west] at (123.500,154.450) {L4};
  \node[font=\fontsize{4.2}{4.6}\selectfont,text=gray!75!black,anchor=east] at (147.000,154.450) {n=72};
  \filldraw[fill=rqheat22,draw=black!10,line width=.20pt] (148.400,148.100) rectangle (159.680,160.800);
  \node[font=\fontsize{6.0}{6.5}\selectfont,text=black!88,anchor=center] at (154.040,154.450) {22};
  \filldraw[fill=rqheat26,draw=black!10,line width=.20pt] (159.680,148.100) rectangle (170.960,160.800);
  \node[font=\fontsize{6.0}{6.5}\selectfont,text=black!88,anchor=center] at (165.320,154.450) {26};
  \filldraw[fill=rqheat25,draw=black!10,line width=.20pt] (170.960,148.100) rectangle (182.240,160.800);
  \node[font=\fontsize{6.0}{6.5}\selectfont,text=black!88,anchor=center] at (176.600,154.450) {25};
  \filldraw[fill=rqheat15,draw=black!10,line width=.20pt] (182.240,148.100) rectangle (193.520,160.800);
  \node[font=\fontsize{6.0}{6.5}\selectfont,text=black!88,anchor=center] at (187.880,154.450) {15};
  \filldraw[fill=rqheat8,draw=black!10,line width=.20pt] (193.520,148.100) rectangle (204.800,160.800);
  \node[font=\fontsize{6.0}{6.5}\selectfont,text=black!88,anchor=center] at (199.160,154.450) {8};
  \filldraw[fill=rqheat0,draw=black!10,line width=.20pt] (204.800,148.100) rectangle (216.080,160.800);
  \node[font=\fontsize{6.0}{6.5}\selectfont,text=black!88,anchor=center] at (210.440,154.450) {0};
  \filldraw[fill=rqheat0,draw=black!10,line width=.20pt] (216.080,148.100) rectangle (227.360,160.800);
  \node[font=\fontsize{6.0}{6.5}\selectfont,text=black!88,anchor=center] at (221.720,154.450) {0};
  \filldraw[fill=rqheat3,draw=black!10,line width=.20pt] (227.360,148.100) rectangle (238.640,160.800);
  \node[font=\fontsize{6.0}{6.5}\selectfont,text=black!88,anchor=center] at (233.000,154.450) {3};
  \node[font=\fontsize{6.0}{6.5}\selectfont,anchor=west] at (123.500,167.150) {L5};
  \node[font=\fontsize{4.2}{4.6}\selectfont,text=gray!75!black,anchor=east] at (147.000,167.150) {n=214};
  \filldraw[fill=rqheat5,draw=black!10,line width=.20pt] (148.400,160.800) rectangle (159.680,173.500);
  \node[font=\fontsize{6.0}{6.5}\selectfont,text=black!88,anchor=center] at (154.040,167.150) {5};
  \filldraw[fill=rqheat12,draw=black!10,line width=.20pt] (159.680,160.800) rectangle (170.960,173.500);
  \node[font=\fontsize{6.0}{6.5}\selectfont,text=black!88,anchor=center] at (165.320,167.150) {12};
  \filldraw[fill=rqheat9,draw=black!10,line width=.20pt] (170.960,160.800) rectangle (182.240,173.500);
  \node[font=\fontsize{6.0}{6.5}\selectfont,text=black!88,anchor=center] at (176.600,167.150) {9};
  \filldraw[fill=rqheat62,draw=black!10,line width=.20pt] (182.240,160.800) rectangle (193.520,173.500);
  \node[font=\fontsize{6.0}{6.5}\selectfont,text=white,anchor=center] at (187.880,167.150) {62};
  \filldraw[fill=rqheat12,draw=black!10,line width=.20pt] (193.520,160.800) rectangle (204.800,173.500);
  \node[font=\fontsize{6.0}{6.5}\selectfont,text=black!88,anchor=center] at (199.160,167.150) {12};
  \filldraw[fill=rqheat0,draw=black!10,line width=.20pt] (204.800,160.800) rectangle (216.080,173.500);
  \node[font=\fontsize{6.0}{6.5}\selectfont,text=black!88,anchor=center] at (210.440,167.150) {0};
  \filldraw[fill=rqheat0,draw=black!10,line width=.20pt] (216.080,160.800) rectangle (227.360,173.500);
  \node[font=\fontsize{6.0}{6.5}\selectfont,text=black!88,anchor=center] at (221.720,167.150) {0};
  \filldraw[fill=rqheat0,draw=black!10,line width=.20pt] (227.360,160.800) rectangle (238.640,173.500);
  \node[font=\fontsize{6.0}{6.5}\selectfont,text=black!88,anchor=center] at (233.000,167.150) {0};
  \node[font=\fontsize{5.7}{6.2}\selectfont,anchor=north] at (154.040,175.500) {S};
  \node[font=\fontsize{5.7}{6.2}\selectfont,anchor=north] at (165.320,175.500) {SG};
  \node[font=\fontsize{5.7}{6.2}\selectfont,anchor=north] at (176.600,175.500) {L};
  \node[font=\fontsize{5.7}{6.2}\selectfont,anchor=north] at (187.880,175.500) {T};
  \node[font=\fontsize{5.7}{6.2}\selectfont,anchor=north] at (199.160,175.500) {V};
  \node[font=\fontsize{5.7}{6.2}\selectfont,anchor=north] at (210.440,175.500) {C};
  \node[font=\fontsize{5.7}{6.2}\selectfont,anchor=north] at (220.720,175.500) {INT};
  \node[font=\fontsize{5.7}{6.2}\selectfont,anchor=north] at (234.000,175.500) {None};
   \foreach \y in {122.700,135.400,148.100,160.800}
    \draw[draw=black,line width=.30pt] (148.400,\y) -- (238.640,\y);
  \draw[draw=black!55,line width=.38pt] (148.400,110.000) rectangle (238.640,173.500);
  \fill[fill=rqlegend0] (54.000,190.000) rectangle (55.426,195.000);
  \fill[fill=rqlegend1] (55.416,190.000) rectangle (56.842,195.000);
  \fill[fill=rqlegend2] (56.832,190.000) rectangle (58.258,195.000);
  \fill[fill=rqlegend3] (58.248,190.000) rectangle (59.673,195.000);
  \fill[fill=rqlegend4] (59.663,190.000) rectangle (61.089,195.000);
  \fill[fill=rqlegend5] (61.079,190.000) rectangle (62.505,195.000);
  \fill[fill=rqlegend6] (62.495,190.000) rectangle (63.921,195.000);
  \fill[fill=rqlegend7] (63.911,190.000) rectangle (65.337,195.000);
  \fill[fill=rqlegend8] (65.327,190.000) rectangle (66.753,195.000);
  \fill[fill=rqlegend9] (66.743,190.000) rectangle (68.168,195.000);
  \fill[fill=rqlegend10] (68.158,190.000) rectangle (69.584,195.000);
  \fill[fill=rqlegend11] (69.574,190.000) rectangle (71.000,195.000);
  \fill[fill=rqlegend12] (70.990,190.000) rectangle (72.416,195.000);
  \fill[fill=rqlegend13] (72.406,190.000) rectangle (73.832,195.000);
  \fill[fill=rqlegend14] (73.822,190.000) rectangle (75.248,195.000);
  \fill[fill=rqlegend15] (75.238,190.000) rectangle (76.663,195.000);
  \fill[fill=rqlegend16] (76.653,190.000) rectangle (78.079,195.000);
  \fill[fill=rqlegend17] (78.069,190.000) rectangle (79.495,195.000);
  \fill[fill=rqlegend18] (79.485,190.000) rectangle (80.911,195.000);
  \fill[fill=rqlegend19] (80.901,190.000) rectangle (82.327,195.000);
  \fill[fill=rqlegend20] (82.317,190.000) rectangle (83.743,195.000);
  \fill[fill=rqlegend21] (83.733,190.000) rectangle (85.159,195.000);
  \fill[fill=rqlegend22] (85.149,190.000) rectangle (86.574,195.000);
  \fill[fill=rqlegend23] (86.564,190.000) rectangle (87.990,195.000);
  \fill[fill=rqlegend24] (87.980,190.000) rectangle (89.406,195.000);
  \fill[fill=rqlegend25] (89.396,190.000) rectangle (90.822,195.000);
  \fill[fill=rqlegend26] (90.812,190.000) rectangle (92.238,195.000);
  \fill[fill=rqlegend27] (92.228,190.000) rectangle (93.654,195.000);
  \fill[fill=rqlegend28] (93.644,190.000) rectangle (95.069,195.000);
  \fill[fill=rqlegend29] (95.059,190.000) rectangle (96.485,195.000);
  \fill[fill=rqlegend30] (96.475,190.000) rectangle (97.901,195.000);
  \fill[fill=rqlegend31] (97.891,190.000) rectangle (99.317,195.000);
  \fill[fill=rqlegend32] (99.307,190.000) rectangle (100.733,195.000);
  \fill[fill=rqlegend33] (100.723,190.000) rectangle (102.149,195.000);
  \fill[fill=rqlegend34] (102.139,190.000) rectangle (103.564,195.000);
  \fill[fill=rqlegend35] (103.554,190.000) rectangle (104.980,195.000);
  \fill[fill=rqlegend36] (104.970,190.000) rectangle (106.396,195.000);
  \fill[fill=rqlegend37] (106.386,190.000) rectangle (107.812,195.000);
  \fill[fill=rqlegend38] (107.802,190.000) rectangle (109.228,195.000);
  \fill[fill=rqlegend39] (109.218,190.000) rectangle (110.644,195.000);
  \fill[fill=rqlegend40] (110.634,190.000) rectangle (112.060,195.000);
  \fill[fill=rqlegend41] (112.050,190.000) rectangle (113.475,195.000);
  \fill[fill=rqlegend42] (113.465,190.000) rectangle (114.891,195.000);
  \fill[fill=rqlegend43] (114.881,190.000) rectangle (116.307,195.000);
  \fill[fill=rqlegend44] (116.297,190.000) rectangle (117.723,195.000);
  \fill[fill=rqlegend45] (117.713,190.000) rectangle (119.139,195.000);
  \fill[fill=rqlegend46] (119.129,190.000) rectangle (120.555,195.000);
  \fill[fill=rqlegend47] (120.545,190.000) rectangle (121.970,195.000);
  \fill[fill=rqlegend48] (121.960,190.000) rectangle (123.386,195.000);
  \fill[fill=rqlegend49] (123.376,190.000) rectangle (124.802,195.000);
  \fill[fill=rqlegend50] (124.792,190.000) rectangle (126.218,195.000);
  \fill[fill=rqlegend51] (126.208,190.000) rectangle (127.634,195.000);
  \fill[fill=rqlegend52] (127.624,190.000) rectangle (129.050,195.000);
  \fill[fill=rqlegend53] (129.040,190.000) rectangle (130.465,195.000);
  \fill[fill=rqlegend54] (130.455,190.000) rectangle (131.881,195.000);
  \fill[fill=rqlegend55] (131.871,190.000) rectangle (133.297,195.000);
  \fill[fill=rqlegend56] (133.287,190.000) rectangle (134.713,195.000);
  \fill[fill=rqlegend57] (134.703,190.000) rectangle (136.129,195.000);
  \fill[fill=rqlegend58] (136.119,190.000) rectangle (137.545,195.000);
  \fill[fill=rqlegend59] (137.535,190.000) rectangle (138.960,195.000);
  \fill[fill=rqlegend60] (138.950,190.000) rectangle (140.376,195.000);
  \fill[fill=rqlegend61] (140.366,190.000) rectangle (141.792,195.000);
  \fill[fill=rqlegend62] (141.782,190.000) rectangle (143.208,195.000);
  \fill[fill=rqlegend63] (143.198,190.000) rectangle (144.624,195.000);
  \fill[fill=rqlegend64] (144.614,190.000) rectangle (146.040,195.000);
  \fill[fill=rqlegend65] (146.030,190.000) rectangle (147.456,195.000);
  \fill[fill=rqlegend66] (147.446,190.000) rectangle (148.871,195.000);
  \fill[fill=rqlegend67] (148.861,190.000) rectangle (150.287,195.000);
  \fill[fill=rqlegend68] (150.277,190.000) rectangle (151.703,195.000);
  \fill[fill=rqlegend69] (151.693,190.000) rectangle (153.119,195.000);
  \fill[fill=rqlegend70] (153.109,190.000) rectangle (154.535,195.000);
  \fill[fill=rqlegend71] (154.525,190.000) rectangle (155.951,195.000);
  \fill[fill=rqlegend72] (155.941,190.000) rectangle (157.366,195.000);
  \fill[fill=rqlegend73] (157.356,190.000) rectangle (158.782,195.000);
  \fill[fill=rqlegend74] (158.772,190.000) rectangle (160.198,195.000);
  \fill[fill=rqlegend75] (160.188,190.000) rectangle (161.614,195.000);
  \fill[fill=rqlegend76] (161.604,190.000) rectangle (163.030,195.000);
  \fill[fill=rqlegend77] (163.020,190.000) rectangle (164.446,195.000);
  \fill[fill=rqlegend78] (164.436,190.000) rectangle (165.861,195.000);
  \fill[fill=rqlegend79] (165.851,190.000) rectangle (167.277,195.000);
  \fill[fill=rqlegend80] (167.267,190.000) rectangle (168.693,195.000);
  \fill[fill=rqlegend81] (168.683,190.000) rectangle (170.109,195.000);
  \fill[fill=rqlegend82] (170.099,190.000) rectangle (171.525,195.000);
  \fill[fill=rqlegend83] (171.515,190.000) rectangle (172.941,195.000);
  \fill[fill=rqlegend84] (172.931,190.000) rectangle (174.357,195.000);
  \fill[fill=rqlegend85] (174.347,190.000) rectangle (175.772,195.000);
  \fill[fill=rqlegend86] (175.762,190.000) rectangle (177.188,195.000);
  \fill[fill=rqlegend87] (177.178,190.000) rectangle (178.604,195.000);
  \fill[fill=rqlegend88] (178.594,190.000) rectangle (180.020,195.000);
  \fill[fill=rqlegend89] (180.010,190.000) rectangle (181.436,195.000);
  \fill[fill=rqlegend90] (181.426,190.000) rectangle (182.852,195.000);
  \fill[fill=rqlegend91] (182.842,190.000) rectangle (184.267,195.000);
  \fill[fill=rqlegend92] (184.257,190.000) rectangle (185.683,195.000);
  \fill[fill=rqlegend93] (185.673,190.000) rectangle (187.099,195.000);
  \fill[fill=rqlegend94] (187.089,190.000) rectangle (188.515,195.000);
  \fill[fill=rqlegend95] (188.505,190.000) rectangle (189.931,195.000);
  \fill[fill=rqlegend96] (189.921,190.000) rectangle (191.347,195.000);
  \fill[fill=rqlegend97] (191.337,190.000) rectangle (192.762,195.000);
  \fill[fill=rqlegend98] (192.752,190.000) rectangle (194.178,195.000);
  \fill[fill=rqlegend99] (194.168,190.000) rectangle (195.594,195.000);
  \fill[fill=rqlegend100] (195.584,190.000) rectangle (197.010,195.000);
  \draw[line width=.35pt,draw=black!70] (54.000,190.000) rectangle (197.000,195.000);
  \node[font=\fontsize{5.8}{6.4}\selectfont,anchor=north] at (54.000,196.400) {0};
  \node[font=\fontsize{5.8}{6.4}\selectfont,anchor=north] at (82.600,196.400) {20};
  \node[font=\fontsize{5.8}{6.4}\selectfont,anchor=north] at (111.200,196.400) {40};
  \node[font=\fontsize{5.8}{6.4}\selectfont,anchor=north] at (139.800,196.400) {60};
  \node[font=\fontsize{5.8}{6.4}\selectfont,anchor=north] at (168.400,196.400) {80};
  \node[font=\fontsize{5.8}{6.4}\selectfont,anchor=north] at (197.000,196.400) {100};
  \node[font=\fontsize{6.6}{7}\selectfont,anchor=center] at (125.5,210.0) {Share of tasks with each salient group (\%)};
\end{tikzpicture}%
}
\caption{Distribution of salient Feature Groups across predicted absolute difficulty levels (L1--L5). Each cell reports the percentage of tasks in that level assigned to the indicated salient-group category; $n$ denotes the number of tasks at each level. Gray rows indicate levels containing no tasks.}
\label{fig:rq2_salient}
\end{figure}
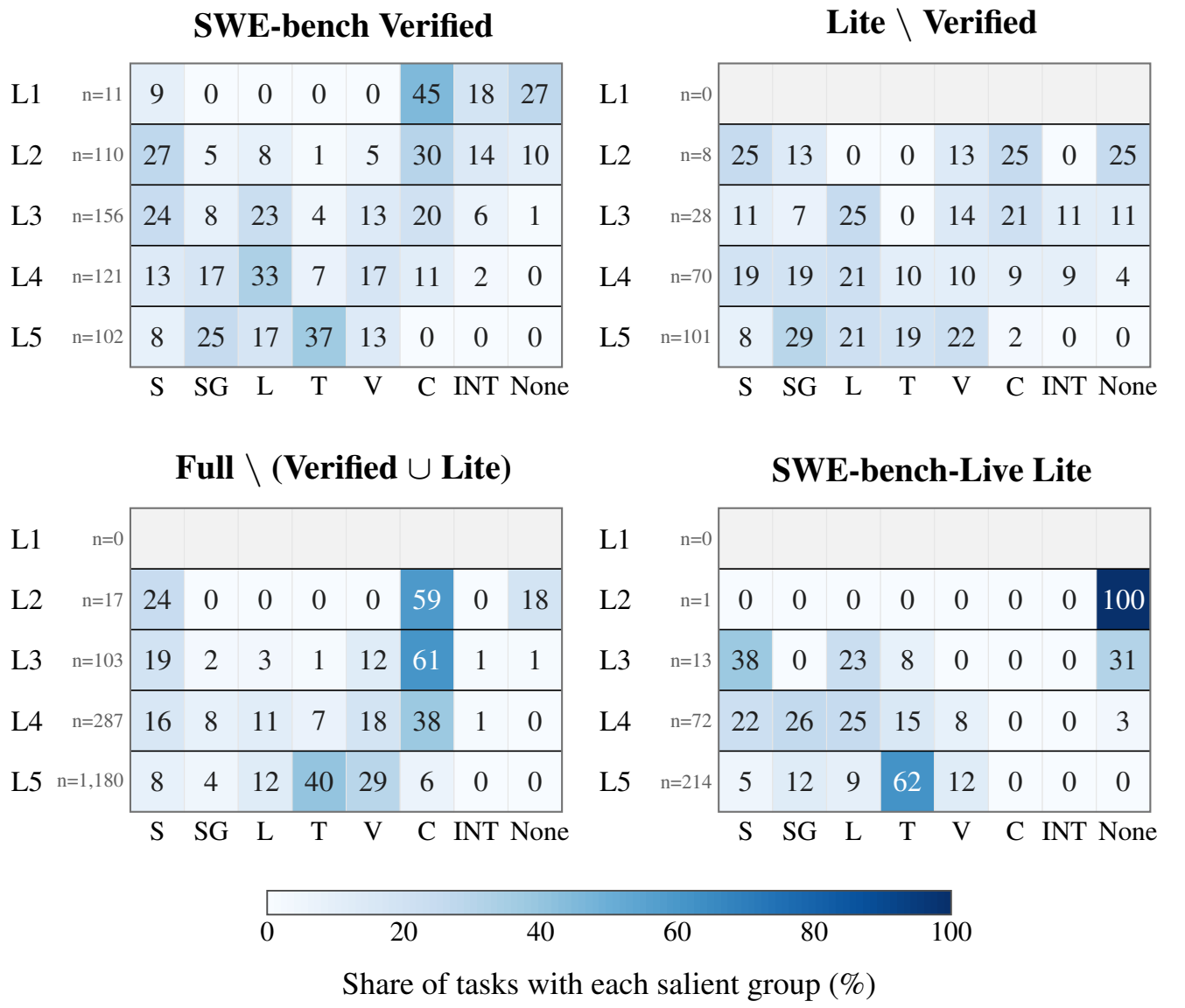


We assign tasks to the predicted difficulty levels (L1--L5) and compare the distribution of their salient Feature Groups (Section~\ref{sec:dare}) within each level.
Figure~\ref{fig:rq2_salient} shows that salient groups varied both across
and within predicted difficulty levels. Across levels, the leading group changed within a benchmark. In SWE-bench Verified, $S$ led at L3 (24\%), $L$ at L4 (33\%), and $T$ at L5 (37\%). Across partitions, $T$ led at L5 in three partitions (37--62\%), whereas $SG$ led in Lite $\setminus$ Verified (29\%). Within levels, salient groups were mixed: the leading group accounted for less than half of the tasks in 13 of the 17 populated levels and at most 62\% in any level with more than one task. Thus, difficulty level alone cannot determine the needed support.

Such task-level variation provides a basis for selecting support according to each task's salient group and evaluating its effectiveness in RQ3. However, L2 contained only one task in SWE-bench-Live Lite and eight in Lite $\setminus$ Verified, so we interpret the distributions at these sparsely populated levels with caution.

\begin{boxA}
\textbf{RQ2 Answer: }Four groups ($T$, $V$, $\mathit{SG}$, and $L$) accounted for 74.4\% of total Feature-Group Importance, but no single group dominated the difficulty model's predictions.
The salient group varied across difficulty levels and benchmark partitions, and even within the same level, so difficulty alone cannot determine the needed support.
\end{boxA}

%% file: sections/rq3.tex
\subsection{RQ3: Does Feature-Group-Guided Support Improve Coding-Agent Repair Success?}
\label{sec:rq3}

RQ1 evaluated \aname{}'s ability to predict repair-task difficulty, while RQ2 identified the Feature Groups contributing most strongly to the predicted difficulty of individual tasks. RQ3 investigates whether support modules selected based on these analyses can improve coding-agent repair success.

\subsubsection{Support Modules}
\label{sec:repair_support}

RQ2 showed that $T$, $V$, $\mathit{SG}$, and $L$ contribute most to difficulty predictions (Table~\ref{tab:rq2_importance}). Among them, we implement modules for $T$, $V$, and $L$, whose repair activities (code change, patch validation, and fault localization) can be supported with existing analysis techniques. We omit $\mathit{SG}$ because we did not identify a sufficiently generalizable support mechanism to directly address this information gap.

\textbf{T-Module.} The T-Module supports large or structurally complex changes. \emph{Change planning} uses repository-level dependency-aware planning~\cite{bairi2024codeplan} to identify follow-up modification locations using static references (Jedi~\cite{jediDocumentation}), type checking (basedpyright~\cite{basedpyrightDocumentation}), inheritance relationships, and co-change history from Git\cite{zimmermann2004rose}. \emph{Completion checking} runs code snippets from the issue and relevant existing tests on the original and modified repository states and reports remaining errors to the agent~\cite{ahmed2025otter,rezaalipour2024fauxpy}.

\textbf{V-Module.} The V-Module assesses the adequacy of the agent's validation. \emph{Change coverage} measures the statement and branch coverage of modified code with coverage.py~\cite{coveragePyDocumentation}, reports untested locations~\cite{mundler2024swtbench,diffCoverDocumentation}, and runs existing tests that exercise modified functions to detect regressions~\cite{chen2026testprune,gligoric2015ekstazi,pytestTestmonDocumentation}. \emph{Validation-gap checks} flag similar expressions left unchanged elsewhere~\cite{poehlmann2013missingfixes}, vary inputs of existing tests~\cite{danglot2019dspot}, and check the consistency of callers and subclasses affected by the change~\cite{ryder2001changeimpact}.

\textbf{L-Module.} The L-Module supports identifying modification locations. It expands initial candidates through call and dependency relationships~\cite{zhang2024autocoderover}, traces executions of agent-written reproduction scripts and, depending on the symptom type, maps error messages to their raising locations and traces faulty values to where they are written~\cite{sridharan2007thinslicing}. These diagnostics direct the agent toward root causes rather than where errors manifest.

All modules use only the issue description and the repository state available to the agent. The complete rules are provided in the replication artifact~\cite{dare_replication}.

\subsubsection{Experimental Setup and Task Selection}

We conduct our experiments using SWE-bench-Live Lite, which contains 300 software repair tasks. To ensure reliable evaluation, we select 118 tasks whose issue descriptions are consistent with the evaluation tests and whose reference patches successfully pass the evaluation. This filtering excludes tasks where failures may result from inconsistencies between issue descriptions and tests or defects in the evaluation environment.

From these tasks, we select three tasks for each of the Transformation (T), Localization (L), and Validation (V) groups, resulting in nine evaluation tasks. Task selection follows two criteria: (1) the corresponding Feature Group is the task’s salient group (Section~\ref{sec:dare}), as identified in the RQ2 analysis, and (2) the task is assigned to predicted difficulty level L4 or L5, corresponding to a final aligned predicted failure rate of at least 0.6. The second criterion focuses the evaluation on tasks predicted to be difficult, where additional support may offer greater room for improvement, while aiming to reduce potential ceiling effects.


\subsubsection{Baseline and Repeated Evaluation}

We use SWE-agent with Claude Sonnet 4.5 as the baseline configuration. At the time of the experiments, this combination achieved the highest performance on the SWE-bench-Live Lite leaderboard among SWE-agent-based configurations that support modular extensions.

All experiments use the same language model and execution and evaluation settings, with access to future Git history containing reference patches blocked. To account for variability across agent executions, each experimental round consists of five baseline runs and five module-assisted runs on the same task. We repeat this procedure for five rounds, resulting in 25 executions per configuration per task and 450 executions across all nine tasks. Repair success is determined by whether the generated patch passes the SWE-bench-Live Lite evaluation tests.

\subsubsection{Statistical Analysis}

We use the two-sided Wilcoxon signed-rank test to evaluate changes in repair performance. For each task, we pair the numbers of successful executions from the baseline and module-assisted configurations across 25 runs, yielding nine paired observations. We set the significance level to 0.05 and additionally report absolute success-rate improvements in percentage points.

\subsubsection{Effect of Feature-Group-Assistant Module}

Table~\ref{tab:rq3_results} presents the repair success counts for the baseline and module-assisted configurations, together with the overall Wilcoxon test result.

\begin{table}[!t]
\centering
\caption{Repair success with and without Assistant Module.}
\label{tab:rq3_results}
\footnotesize
\setlength{\tabcolsep}{6pt}
\renewcommand{\arraystretch}{1.12}
\begin{tabular}{@{}c@{\hspace{5pt}}ccc@{}}
\toprule
Feature Group & Task & Baseline & With Module \\
\midrule
L & matplotlib-29721 & 8/25 & \textbf{21/25} \\
L & reflex-4371 & 16/25 & \textbf{19/25} \\
L & keras-20443 & 16/25 & \textbf{23/25} \\
\midrule
T & pybamm-4865 & 2/25 & \textbf{13/25} \\
T & pylint-10089 & 18/25 & \textbf{19/25} \\
T & starlette-2812 & 18/25 & \textbf{22/25} \\
\midrule
V & pybamm-4644 & 19/25 & \textbf{25/25} \\
V & conan-17967 & 18/25 & \textbf{20/25} \\
V & haystack-8489 & 8/25 & \textbf{17/25} \\
\midrule

\textbf{Total} & & \textbf{123/225} & \textbf{179/225} \\
\midrule
\multicolumn{2}{@{}l}{\textbf{Wilcoxon $p$-value}}
& \multicolumn{2}{c}{\textbf{.0039}} \\
\bottomrule
\end{tabular}
\end{table}

In Table~\ref{tab:rq3_results}, each row represents an evaluation task. The Baseline column reports the number of successful executions using SWE-agent with Claude Sonnet 4.5, while the With Module column reports the corresponding results after applying the Feature Group-specific support module. Each value indicates the number of successful executions out of 25 runs.

The L, T, and V modules improve repair success rates by 30.7, 21.3, and 22.7 percentage points, respectively. The overall success rate increases from 54.7\% (123/225) to 79.6\% (179/225), an improvement of 24.9 percentage points, with improvements observed across all nine tasks. The Wilcoxon signed-rank test yields $p=.0039$, indicating a statistically significant difference at the 0.05 significance level.
    
\begin{boxA}
\textbf{RQ3 Answer: }\aname{} shows that applying support modules corresponding to the most salient Feature Groups can significantly improve repair success ($p=.0039$) on the selected tasks.
\end{boxA}

%% file: sections/discussion.tex
\section{Discussion}\label{sec:discussion}
\subsection{Implications for Agentic Software Engineering}

Hassan et al.~\cite{hassan2026agentic} emphasized the importance of systematic workflows and structured tool support for advancing Agentic Software Engineering. In line with this direction, \aname{} demonstrates how Feature Group contributions to difficulty predictions can inform practical agent support. In RQ3, support modules targeting Localization, Transformation, and Validation improved repair success on selected tasks. These findings suggest that task-specific difficulty analysis can provide practical guidance for agent assistance and motivate further investigation across diverse tasks and agent configurations.

\subsection{Threats to Validity}\label{sec:threats-to-validity}
Difficulty is defined by failure rates, which depend on the evaluated agents; strong cross-population agreement and shared-agent scale alignment mitigate this threat. Reference patches are never given to agents, but patch-derived features indirectly inform support selection, limited to choosing a coarse Feature Group. RQ3 tasks predate the model's training cutoff and may be memorized; however, baseline and module-assisted runs use the same model and tasks, so contamination cannot explain their difference. RQ3 compares matched support only with no support and uses one agent configuration and nine tasks, so part of the gain may come from the modules themselves. All tasks are Python issues with reference patches; broader validation is needed to generalize these results.

%% file: sections/conclusion.tex
\section{Conclusion}\label{sec:conclusion}
We presented \aname{}, a framework that predicts repair-task difficulty from static benchmark artifacts, explains each prediction through seven
Repair-Task Feature Groups, and uses the most salient group to select
task-specific agent support. The difficulty model consistently predicted
the relative difficulty of tasks on unseen benchmark partitions. Matching
support to each task's salient group raised a coding agent's repair success
from 54.7\% to 79.6\% on nine high-difficulty tasks (${p=.0039}$). These results show that interpreting
difficulty predictions can guide how agents are assisted. We plan to generalize this predict--interpret--support process to other agentic tasks.